\documentclass[aps,pra,twocolumn,amsmath,amssymb,notitlepage,floatfix,10pt,longbibliography,superscriptaddress]{revtex4-2}

\usepackage{bbm}
\usepackage{times}
\usepackage{color}
\usepackage{amsthm}
\usepackage[caption=false]{subfig}
\usepackage{multirow}
\usepackage[normalem]{ulem}
\usepackage{amsfonts,amssymb,dsfont,bm}
\usepackage{graphicx}
\usepackage{soul}
\usepackage{hyperref}
\usepackage{lipsum}
\hypersetup{
    colorlinks=true,       
    linkcolor=red,          
  citecolor=magenta,        
    filecolor=magenta,      
    urlcolor=cyan,          
    runcolor=cyan
}
  
\usepackage[acronym,shortcuts]{glossaries}

\renewcommand\Re{\operatorname{Re}}
\renewcommand\Im{\operatorname{Im}}

\DeclareMathOperator{\Tr}{Tr}

\newcommand{\sx}[1]{X_{#1}}
\newcommand{\sy}[1]{Y_{#1}}
\newcommand{\sz}[1]{Z_{#1}}

\newacronym{ARMA}{ARMA}{autoregressive moving-average}
\newacronym{SchWARMA}{SchWARMA}{Schr\"{o}dinger Wave Autoregressive Moving Average}
\newacronym{QNS}{QNS}{quantum noise spectroscopy}
\newacronym{PSD}{PSD}{power spectral density}
\newacronym{FF}{FF}{filter functions}
\newacronym{fttps}{FTTPS}{Fixed Total Time Pulse Sequences}

\begin{document} 

\glsdisablehyper

\title{Assessing Spatiotemporally Correlated Noise in Superconducting Qubits via Pulse-Based Quantum Noise Spectroscopy}
\author{Mayra Amezcua}
\author{Leigh Norris}
\author{Tom Gilliss}
\author{Ryan Sitler}
\author{James Shackford}
\author{Gregory Quiroz}
\author{Kevin Schultz}
\affiliation{Johns Hopkins Applied Physics Laboratory, Laurel, MD 20723, USA}

\date{\today}

\begin{abstract}
Spatiotemporally correlated errors are widespread in quantum devices and are particularly adversarial to error correcting schemes.  To characterize these errors, we propose and validate a nonparametric quantum noise spectroscopy (QNS) protocol to estimate both spectra and static errors associated with spatiotemporally correlated dephasing noise and fluctuating quantum crosstalk on two qubits. Our scheme reconstructs the real and imaginary components of the two-qubit cross-spectrum  by using fixed total time pulse sequences and single qubit and joint two-qubit measurements to separately resolve spatially correlated noise processes. We benchmark our protocol by reconstructing the spectra of spatiotemporally correlated noise processes engineered via the Schr\"{o}dinger Wave Autoregressive Moving Average technique, emulating dephasing errors. Furthermore, we show that the protocol can outperform existing comb-based QNS protocols. Our results demonstrate the utility of our protocol in characterizing spatiotemporally correlated noise and quantum crosstalk in a multi-qubit device for potential use in noise-adapted control or error protection schemes. 
\end{abstract}

\maketitle

\section{Introduction}
Quantum error correcting codes (QECCs) are key components for achieving large-scale fault-tolerant quantum computing~\cite{shor1996fault,gottesman1998ftqc}. A primary challenge in modern quantum computing is bridging the gap between the error models assumed by QECCs and the observed dynamics of real devices. Typical noise models for QECCs assume noise channels with spatial and temporal locality, however, this assumption is commonly violated in today's quantum devices. For example, high-energy particles have been known to cause correlated errors that spread through superconducting qubit devices~\cite{wilen2021correlated,mcewen2022resolving}. Moreover, the prevalence of temporally correlated dephasing and control noise~\cite{bylander2011:fnoise, yan2013:fnoise, burnett2014:fnoise, frey2017application:ions, chan2018:spin, Buterakos2018semiconductor, burnett2019:fnoise, struck2020:fnoise, connors2022charge, Fang2022trapped-ion}, as well as parasitic crosstalk errors~\cite{PhysRevLett.131.210802, heinz2021xtalk, Heinz2022spinqubit, zhao2022xtalk, fors2024comprehensiveexplanationzzcoupling}, has been extensively documented across a range of quantum computing platforms. The impact of these noise sources has been observed in a variety of error protection protocols designed to suppress, correct, or mitigate errors~\cite{klessequantum,ng2009fault, premakumar1812error, ding2020xtalk, PhysRevLett.131.210802,quiroz2024dynamically:dfs, schultz2022impact:zne, seif2024suppressing, tripathi2022xtalk, evert2024syncopateddynamicaldecouplingsuppressing, vezvaee2025demonstration:lq-xtalk, brown2025dd, hickman2025crosstalkrobustdynamicaldecouplingbipartitetopology}. In order for effective error protection to be realized---whether it be via below-threshold QECCs or other approaches---it is necessary for the protocols to be informed by accurate noise models.

Noise characterization is a crucial component of error model development. It can facilitate predictive dynamics modeling and assist in targeted error protection protocol design. A common approach is to utilize specific types of circuit sampling to reveal noise characteristics; this includes variants of randomized benchmarking~\cite{emerson2005scalable,magesan2012efficient,gambetta2012characterization,helsen2019new,erhard2019characterizing:rb,helsen2022general}, gate set tomography~\cite{nielsen2021gate}, Pauli noise estimation~\cite{flammia2020efficient}, and average circuit eigenvalue sampling~\cite{flammia2021averaged}. Relying on randomization for noise tailoring, these methods enable specific assumptions to be made about the noise model. However, sufficient sampling is required to ensure effective noise tailoring, which can be costly as the system size increases~\cite{proctor2019rb, helsen2019rb, proctor2022mb}. An important assumption typically made by these methods is that the noise is Markovian, i.e., it does not exhibit temporal correlations. As such, traditionally, they do not reveal information about noise correlations.

Quantum noise spectroscopy (QNS) techniques, on the other hand, are specifically designed to characterize spatiotemporally correlated noise. They use the quantum system as a dynamical sensor to characterize the spectral properties of noise processes. QNS typically involves applying a set of control sequences to a quantum system in order to tune the system's frequency response, i.e., filter function (FF). The connection between the FF and the spectral properties of the noise via the filter function formalism~\cite{cywinski2008:fff, green2013arbitrary, pazsilva2014transferfunc, pazsilva2017:2qqns} facilitates the estimation of the noise polyspectra.

Pulse-based and continuous control single-qubit \acrshort{QNS} have been demonstrated on several qubit platforms: superconducting qubits~\cite{bylander2011:fnoise,yan2013:fnoise,sung2019nong:qns,murphy2021universal,PhysRevLett.131.210802,oda2024sparse}, trapped-ions~\cite{frey2017application:ions}, nitrogen-vacancy centers~\cite{romach2015spectroscopy:nv}, and semiconductor quantum dots~\cite{yondea2018:spin,chan2018:spin}. These protocols have been extended to multi-qubit systems~\cite{rivas2015quantifying, pazsilva2017:2qqns,szankowski2016:2qqns,boter2020dots:2qqns,lupke2020experiment:2qqns,yoneda2023noise:2qqns,rojasarias2023nnn:2qqns} to identify spatially correlated errors and used to characterize Gaussian and non-Gaussian noise in classical and quantum regimes~\cite{pazsilva2017:2qqns, norris2016qubit, sung2019nong:qns}. Pulse-based techniques require design and implementation of long pulse sequences which are limited by coherence and coherent errors. To probe a wide frequency range, continuous control protocols require a calibrated pulse per frequency of interest, increasing calibration overhead which is susceptible to drift. Importantly, a majority of the approaches do not support characterization of qubit-qubit interactions---commonly referred to as quantum crosstalk---encountered on a wide range of devices~\cite{Buterakos2018semiconductor, ashsak2020xt, kandala2021xt, zhao2022xtalk, Heinz2022spinqubit, throckmorton2022xt}.

Here, we introduce a comprehensive pulse-based, two-qubit protocol for characterizing spatiotemporally correlated dephasing noise across pairs of qubits. The method estimates the first and second moments of the noise, providing information about the static components and self- and cross-spectra of the single-qubit noise. The static components induce errant $z$-rotations that are used to estimate coherent dephasing and detuning errors. An important feature of this protocol is that it treats DC spectral contributions separately, and therefore the broader spectral reconstructions are robust to coherent dephasing and detuning errors. In addition, the protocol affords the unique feature of being sensitive to quantum crosstalk, and thus, can be used to reveal information about the static and spectral characteristics of the joint, two-qubit noise.

A key aspect of the protocol is its use of \acrfull{fttps}~\cite{murphy2021universal,PhysRevLett.131.210802}. These pulse sequences improve upon previous approaches by requiring only single-qubit gates and possessing a constant duration that can be tuned based on the characteristic timescale of the qubit, e.g., $T_1$ relaxation times. Moreover, the \acrshort{fttps} FFs are well-localized in frequency space. This results in better-conditioned reconstruction matrices than previously utilized pulse sequences~\cite{murphy2021universal}; thus, improving reconstruction accuracy.

The protocol is validated through numerical studies and experiments. In addition to supporting the efficacy of the protocol, the numerical simulations assist in characterizing the estimation error and its relation to the magnitude of the noise parameters for the noise spectrum. Experiments are performed on a fixed-frequency, fixed-coupler, superconducting device, where the protocol is subjected to native and engineered noise environments. The latter involves injecting spatiotemporally correlated noise via \acrfull{SchWARMA} models~\cite{schultz2020schwarma}. We show that our protocol provides accurate reconstructions of the injected noise spectrum, exceeding the performance of typical frequency comb-based techniques specifically for sharp spectral features. Overall, our protocol offers a practical approach to pulse-based multi-qubit QNS that affords enhanced reconstruction accuracy with less complexity in sequence and pulse overheads than existing techniques.

The structure of the manuscript is as follows. In Section~\ref{sec:NoiseCtrl}, we introduce the system model, describing a driven two-qubit system subject to correlated noise and quantum crosstalk. We show how the qubit dynamics are related to the overlap of the control FFs and noise power spectra. In Section~\ref{sec:Protocol}, we describe the steps of the \acrshort{QNS} protocol and outline our approach to estimating the static noise terms and power spectra. We examine the efficacy of the protocol via numerical simulations in Sec.~\ref{sec:Simulations}. The simulations are complemented with experimental investigations in Sec.~\ref{sec:Experiments}, where we examine injected and native noise environments. We include a discussion on how measurement error mitigation techniques impact the noise reconstruction and compare our protocol to the frequency comb approach. Finally, Section~\ref{sec:Conclusion} summarizes our main results and discusses future work.

\section{Noise and Control Setting}\label{sec:NoiseCtrl}
We consider a system of two qubits subject to spatiotemporally correlated dephasing noise, quantum crosstalk, and control. In the rotating frame, the system dynamics are generated by the Hamiltonian $H(t)=H_\text{int}(t)+H_\text{ctrl}(t)$. Here, the control Hamiltonian, $H_\text{ctrl}(t)$, describes the ideal controlled dynamics and the noise Hamiltonian, $H_\text{int}(t)$, captures departures from the ideal dynamics due to spatiotemporally correlated dephasing and multi-qubit $ZZ$ crosstalk.
The noise Hamiltonian takes the form
\begin{align}
H_\text{int}(t)=\sum_{n=1}^{2}[\Delta_n+\beta_{n}(t)]\sz{n} + [J+\beta_{12}(t)]\sz{1}\sz{2},
\label{eq:H-int}
\end{align}
where $\{\sx{n},\sy{n},\sz{n}\}$ denote the Pauli operators on the $n$th qubit. The stochastic process $\beta_{n}(t)$ describes dephasing noise acting on qubit $n$ and $\Delta_{n}$ represents a static error source, e.g., qubit frequency detuning. The qubits interact via $ZZ$-crosstalk with a mean strength $J$ and fluctuations captured by the stochastic process $\beta_{12}(t)$. We treat $\beta_{1}(t)$, $\beta_{2}(t)$ and $\beta_{12}(t)$ as stationary, zero-mean, and Gaussian. The crosstalk noise $\beta_{12}(t)$ is assumed to be statistically independent of $\beta_{1}(t)$ and $\beta_{2}(t)$, but we allow for the possibility of spatiotemporal correlations between $\beta_{1}(t)$ and $\beta_{2}(t)$. Our noise model differs from those considered in previous multi-qubit dynamical decoupling-based noise spectroscopy protocols~\cite{szankowski2016:2qqns,pazsilva2017:2qqns} due to the presence of the static terms $\Delta_1$, $\Delta_2$, and $J$, as well as time-correlated crosstalk $\beta_{12}(t)$. Control on qubit $n$ is generated by varying the amplitude $\Omega_{n}(t)$ and phase $\phi_{n}(t)$ of an external driving field, as described by the control Hamiltonian
% Control Hamiltonian
\begin{align}
    H_\text{ctrl}(t) = \sum_{n\in\{1,2\}}\Omega_{n}(t)[\cos{\phi_{n}(t)}\sx{n}+\sin{\phi_{n}(t)}\sy{n}].
\end{align}
We assume that qubits 1 and 2 are independently controllable, allowing for the possibility that $\Omega_{1}(t)\neq\Omega_{2}(t)$ and $\phi_{1}(t)\neq\phi_{2}(t)$.

To separate the dynamical contributions of the noise and control, it is useful to work in the toggling frame or interaction picture associated with $H_\text{ctrl}(t)$. We consider control consisting of near-instantaneous $\pi$-pulses about $\sx{n}$ or $\sy{n}$, in which case the toggling frame Hamiltonian is given by
\begin{align}
   & \tilde{H}(t) = U_\text{ctrl}(t)^{\dagger}H_\text{int}(t)U_\text{ctrl}(t) \label{eq:Htog}\\ 
    &= \sum_{n=1}^{2}y_{n}(t)[\Delta_n+\beta_{n}(t)]\sz{n} + y_{1}(t)y_{2}(t)[J+\beta_{12}(t)]\sz{1}\sz{2}, \notag
\end{align}
where $U_\text{ctrl}(t)=\mathcal{T}_{+}e^{-i\int_{0}^{t}ds H_\text{ctrl}(s)}$ is the control propagator and $\mathcal{T}_{+}$ is the time-ordering operator. The control switching function associated with qubit $n$, $y_{n}(t)=\Tr\left[\sz{n}U_\text{ctrl}(t)^{\dagger}\sz{n}U_\text{ctrl}(t)\right]$, changes sign between $\pm1$ each time a $\pi$-pulse is applied. Note that this expression is exact only in the idealized limit of instantaneous $\pi$-pulses. In Appendix \ref{app:pulsewidth_errors}, we examine the effects of finite pulse widths and how they can be suppressed. Evolution of the system in the toggling frame is related to evolution in the rotating frame via $U(t)=U_\text{ctrl}(t)\tilde{U}(t)$, where $U(t)=\mathcal{T}_{+}e^{-i\int_{0}^{t}ds H(s)}$ is the rotating-frame propagator and $\tilde{U}(t)=\mathcal{T}_{+}e^{-i\int_{0}^{t}ds \tilde{H}(s)}$ is the toggling frame propagator. Our QNS protocol employs control sequences with even numbers of $\pi$-pulses over a time duration $t=T$, such that $U_\text{ctrl}(T)=I$, ensuring that system evolution in the rotating frame and toggling frame are equivalent.

\subsection{Dynamics}
Following the approach in Ref.~\cite{pazsilva2017:2qqns}, we use a cumulant expansion to determine the expected values of observables on one or both qubits. After allowing the qubits to evolve from initial state $\rho_0$ under spatiotemporally correlated single-qubit noise, crosstalk, and control for a time $T$, the expected value of an invertible observable $O$ is given by
\begin{align}
\mathbb{E}(O(T))=&\,\text{Tr}[\big\langle O^{-1}\tilde{U}(T)^\dag O \tilde{U}(T) \big\rangle \rho_0 O]\notag\\
=&\,\text{Tr}[e^{\sum_{k=1}^\infty (-i)^k \frac{\mathcal{C}_O^{(k)}(T)}{k!}} \rho_0 O].\label{eq:expO}
\end{align}
Here, $\langle\cdot\rangle$ denotes an ensemble average over the noise processes and $\mathcal{C}_O^{(k)}(T)$ is the $k$th order cumulant, which depends on the observable $O$ as well as the noise terms. Since the dynamics generated by the toggling frame Hamiltonian are purely dephasing and we have assumed that the noise processes are Gaussian, cumulants of order $k>2$ vanish. We need only consider the cumulants of order $k\leq 2$. 

The leading order $k=1$ and $k=2$ cumulants generate different dynamical effects. The first $(k=1)$ cumulant represents the dynamical contribution of the static noise terms,
\begin{widetext}
\begin{align}
\mathcal{C}_O^{(1)}(T)=\sum_{n=1}^2\int_0^Tdt\, y_n(t)\Delta_n(Z_n-O^{-1}Z_nO)
+\int_0^T dt\, y_1(t)y_2(t)J(Z_1Z_2-O^{-1}Z_1Z_2O), \label{eq:C1}
\end{align}
while the second $(k=2)$ cumulant captures the dynamics induced by the time-dependent noise terms,
\begin{align}\notag
\frac{\mathcal{C}_O^{(2)}(T)}{2!}=&\,\frac{1}{2}\int_0^Tdt_1\int_0^Tdt_2\sum_{n,n'=1}^2y_n(t_1)y_{n'}(t_2)\langle\beta_n(t_1)\beta_{n'}(t_2)\rangle(Z_nZ_{n'}+O^{-1}Z_nZ_{n'}O-O^{-1}Z_nOZ_{n'}-O^{-1}Z_{n'}OZ_{n})\\\label{eq:C2}
&+\int_0^Tdt_1\int_0^Tdt_2y_{12}(t_1)y_{12}(t_2)\langle\beta_{12}(t)\beta_{12}(t)\rangle(I-O^{-1}Z_1Z_2OZ_1Z_2).
\end{align}
\end{widetext}
Here, $y_{12}(t)\equiv y_1(t)y_2(t)$. Time-dependent noise enters the second cumulant through the correlation functions $\langle\beta_n(t_1)\beta_{n'}(t_2)\rangle$ and $\langle\beta_{12}(t_1)\beta_{12}(t_2)\rangle$. 
Stationarity implies the correlation functions depend only on the lag time $\tau\equiv t_1-t_2$, giving them functional forms $\langle\beta_n(t_1)\beta_{n'}(t_2)\rangle=\gamma_{n,n'}(\tau) $ and $\langle\beta_{12}(t_1)\beta_{12}(t_2)\rangle=\gamma_{12,12}(\tau)$. 
Note that in the temporally uncorrelated, white noise limit $\gamma_{n,n'}(\tau),\, \gamma_{12,12}(\tau) \propto\delta(\tau)$, implying that $\mathbb{E}(O(T))$ decays exponentially with $T$. For the stationary, temporally correlated noise that we consider, $\gamma_{n,n'}(\tau)$ and $\gamma_{12,12}(\tau)$ have a more general dependence on the lag time. Spatial correlations between $\beta_1(t)$ and $\beta_2(t)$ imply that cross correlations are generally nonzero, i.e., $\gamma_{1,2}(\tau)=\gamma_{2,1}(\tau)\neq 0$. For the spatially uncorrelated case, in contrast, $\gamma_{1,2}(\tau)=\gamma_{2,1}(\tau)\propto\langle\beta_{1}\rangle \langle\beta_{2}\rangle\delta_{1,2}$, which is zero because $\beta_{n}$ is zero-mean.

\subsection{The Frequency Domain}

To relate the observable expectation values to the spectra, we transform \eqref{eq:C2} into the frequency domain:
\begin{widetext}
\begin{align}\notag
\frac{\mathcal{C}_O^{(2)}(T)}{2!}=&\,\frac{1}{2\pi}\int_0^\infty d\omega\sum_{n,n'=1}^2 G_{n,n'}(\omega,T)S_{n,n'}(\omega)(Z_nZ_{n'}+O^{-1}Z_nZ_{n'}O-O^{-1}Z_nOZ_{n'}-O^{-1}Z_{n'}OZ_{n})\\&+
\frac{1}{\pi}\int_0^\infty d\omega\, G_{12,12}(\omega,T)S_{12,12}(\omega)(I-O^{-1}Z_1Z_2OZ_1Z_2).\label{eq:C2Frequency}
\end{align}
\end{widetext}
In the frequency domain, the effects of the control are described by the \acrshort{FF}
\begin{align}\label{eq:FFdefinition}
G_{m,m'}(\omega,T)=\bigg[\int_0^Tdt\,y_m(t)e^{i\omega t}\bigg] \bigg[\int_0^Tdt\,y_{m'}(t)e^{-i\omega t}\bigg],
\end{align}
where $m,m'\in\{1,2,12\}$. The influence of the noise is captured by the noise power spectral densities (PSDs)
\begin{align}
&S_{n,n'}(\omega)=\int_{-\infty}^\infty d\tau\, e^{-i\omega\tau}\gamma_{n,n'}(\tau),\\
&S_{12,12}(\omega)=\int_{-\infty}^\infty d\tau\, e^{-i\omega\tau}\gamma_{12,12}(\tau).
\end{align}
Observe that, in \eqref{eq:C2Frequency}, the dynamical contribution of the noise depends on the frequency-domain overlap between the spectra and FFs. To suppress noise, it is desirable to select control that produces FFs with minimal spectral overlap. For QNS, on the other hand, the FFs should be shaped in a way that enables us to gain information about the spectra through measurements on the qubits. 

The goal of our QNS procedure is to estimate the spectra associated with both noise on the individual qubits and noise that is correlated between the qubits. Broadly, we can divide the spectra into two categories.
The spectra $S_{1,1}(\omega)$, $S_{2,2}(\omega)$, and $S_{12,12}(\omega)$, which depend on the autocorrelation of a single noise process, are known as self-spectra.
The self-spectra $S_{1,1}(\omega)$ and $S_{2,2}(\omega)$ characterize temporally correlated noise acting on qubits $1$ and $2$, respectively, while $S_{12,12}(\omega)$ characterizes temporally correlated fluctuations in the crosstalk between the qubits. The spectral properties of spatiotemporally correlated noise are captured by the cross-spectra, $S_{1,2}(\omega)$ and $S_{2,1}(\omega)$, which are nonzero only when correlations between $\beta_1(t)$ and $\beta_2(t)$ are present. The crosstalk is assumed to be independent of $\beta_{1}(t)$ and $\beta_{2}(t)$, so cross-correlations between the crosstalk and single-qubit dephasing noise are not present. The cross-spectra are generally complex, with $S_{1,2}(\omega)=S_{2,1}(\omega)^*$ for the classical noise considered here. The self-spectra, in contrast, are always real. In the following section, we will outline a protocol for estimating $S_{1,1}(\omega)$, $S_{2,2}(\omega)$, $S_{12,12}(\omega)$, $\text{Re}[S_{1,2}(\omega)]$, and $\text{Im}[S_{1,2}(\omega)]$.

\section{Protocol}\label{sec:Protocol}
\acrshort{QNS} protocols typically involve: (1) preparing a quantum system in an initial state that is sensitive to the target noise; (2) allowing the system to evolve under noise and control; (3) measuring an observable whose expectation depends on the noisy dynamics; and (4) repeating this process for different state preparations, control sequences, and observables to obtain the data required to estimate the relevant noise spectra over a range of frequencies. Below, we outline an approach to estimate the self- and cross-spectra for the two-qubit system described above, including the necessary observables, state preparations, and control sequences.

\subsection{Observable-Specific Cumulants}
To estimate the spectra, we rely on two types of observables: single qubit observables of the form $O_n$ and two-qubit observables of the form $O_{1}O_{2}$, where $O_n\in\{X_{n},Y_{n}\}$ for $n\in\{1,2\}$.  From Eqs.~\eqref{eq:expO}-\eqref{eq:C2}, the expected value of a single-qubit observable $O_n$, $\mathbb{E}[O_n(T)]$, depends on the first and second cumulants, which take the explicit form
\begin{align}
 \mathcal{C}_{O_n}^{(1)}(T) = &\,\Delta_{n}\left[2 \int_{0}^{T}dt\,y_{n}(t)\right]Z_n\notag
\\ &+J\left[2 \int_{0}^{T}dt\,y_{12}(t)\right]Z_1Z_2\notag\\
 \equiv&\, \theta_{n}(T)\,Z_n+\theta_{12}(T)\,Z_1Z_2,\label{eq:C1Obs1}
 \end{align}
 \begin{align}
 \frac{\mathcal{C}_{O_n}^{(2)}(T)}{2!} =&\, \frac{2}{\pi}\int_0^\infty d\omega\,G_{n,n}(\omega,T)S_{n,n}(\omega)I\notag\\
 &\,+\frac{2}{\pi}\int_0^\infty d\omega\,G_{12,12}(\omega,T)S_{12,12}(\omega)I\notag\\
\equiv &\,\chi_{n,n;12,12}(T)\,I.\label{eq:1qDecay}
\end{align}
Here, the dynamical contributions of the static noise and crosstalk, which we denote by the rotation angles $\theta_{n}(T)$ and $\theta_{12}(T)$, enter $\mathbb{E}[O_n(T)]$ through the first cumulant. The dynamical contributions of the spectra $S_{n,n}(\omega)$ and $S_{12,12}(\omega)$, which we denote by the decay rate $\chi_{n,n;12,12}(T)$, enter through the second cumulant.
The expected value of a two-qubit observable $O_{1}O_{2}$ has a similar dependence on the first and second cumulants,
\begin{align}
\mathcal{C}_{O_1O_2}^{(1)}(T) =&\, \Delta_{1}\left[2 \int_{0}^{T}dt\,y_{1}(t)\right]Z_1 \notag\\
&+\Delta_{2}\left[2 \int_{0}^{T}dt\,y_{2}(t)\right]Z_2 \notag\\
\equiv&\, \theta_{1}(T)\,Z_1+\theta_{2}(T)\,Z_2,\label{eq:C1Obs2}
\end{align}
\begin{align}
\frac{\mathcal{C}^{(2)}_{O_{1}O_{2}}(T)}{2!} = &\,\frac{2}{\pi}\int_{0}^{\infty}\!\!\!d\omega\,G_{1,1}(\omega,T)S_{1,1}(\omega)I\notag\\ 
&+ \frac{2}{\pi}\int_{0}^{\infty}\!\!\!d\omega\,G_{2,2}(\omega,T)S_{2,2}(\omega)I\notag\\
    &+ \frac{4}{\pi}\int_{0}^{\infty}\!\!\!d\omega\Re\!\big[G_{1,2}(\omega,T)\big]\!\Re\!\big[S_{1,2}(\omega)\big]\sz{1}\sz{2}\notag\\
    &-\frac{4}{\pi}\int_{0}^{\infty}\!\!\!d\omega\Im\!\big[G_{1,2}(\omega,T)\big]\!\Im\!\big[S_{1,2}(\omega)\big]\sz{1}\sz{2}\notag\\
    \equiv&\,\chi_{1,1;2,2}(T)I+\chi_{1,2}(T)\sz{1}\sz{2}.\label{eq:2qDecay}
\end{align}
The first cumulant again depends on the dynamical contributions of the static noise components, which we denote by $\theta_{1}(T)$ and $\theta_{2}(T)$. The spectra $S_{1,1}(\omega)$, $S_{2,2}(\omega)$, $\Re[S_{1,2}(\omega)]$, and $\Im[S_{1,2}(\omega)]$ enter through the second cumulant. In the second cumulant, we denote the dynamical contribution of $S_{1,1}(\omega)$ and $S_{2,2}(\omega)$ by the decay rate $\chi_{1,1;2,2}(T)$ and the dynamical contribution of $\Re[S_{1,2}(\omega)]$ and $\Im[S_{1,2}(\omega)]$ by the decay rate $\chi_{1,2}(T)$. Note that the crosstalk spectrum does not contribute to the second cumulant for the two-qubit observable case. This arises from the commutativity between the two-local observable and nose operators described in the second term of Eq.~\eqref{eq:C2Frequency}.

\subsection{Extracting Static and Dynamic Noise Parameters}

The cumulants above show that, in addition to depending on the spectra we wish to estimate, the expected values depend on $\Delta_1$, $\Delta_2$, and $J$. Taking an approach similar to Ref. \cite{PazSilva2016}, we can isolate the spectra-dependent decay rates by combining measurements of Pauli observables with specific initial state preparations. Let $\mathbb{E}_{P_1P_2}[O(T)]$ denote the expected value of operator $O$ at time $T$ when qubit 1 is initially prepared in a +1 eigenstate of the Pauli operator $P_1$ and qubit 2 is initially prepared in a +1 eigenstate of the Pauli operator $P_2$. Using Eqs.~\eqref{eq:expO} and \eqref{eq:2qDecay}, we can isolate the decay rates $\chi_{1,1;12,12}(T)$ and $\chi_{2,2;12,12}(T)$ from four single-qubit expected values:
\begin{align}
&\mathbb{E}_{XZ}[X_1(T)]=e^{-\chi_{1,1;12,12}(T)}\cos[\theta_{1}(T)+\theta_{12}(T)],\\
&\mathbb{E}_{XZ}[Y_1(T)]=e^{-\chi_{1,1;12,12}(T)}\sin[\theta_{1}(T)+\theta_{12}(T)],\\
&\mathbb{E}_{ZX}[X_2(T)]=e^{-\chi_{2,2;12,12}(T)}\cos[\theta_{2}(T)+\theta_{12}(T)],\\
&\mathbb{E}_{ZX}[Y_2(T)]=e^{-\chi_{2,2;12,12}(T)}\sin[\theta_{2}(T)+\theta_{12}(T)].
\end{align}
In terms of the single-qubit expected values, the decay rates $\chi_{1,1;12,12}(T)$ and $\chi_{2,2;12,12}(T)$ are given by
\begin{align}
&\chi_{1,1;12,12}(T)=-\frac{\text{ln}\{\mathbb{E}_{XZ}[X_1(T)]^2+\mathbb{E}_{XZ}[Y_1(T)]^2\}}{2},\label{eq:chi11JJ}\\
&\chi_{2,2;12,12}(T)=-\frac{\text{ln}\{\mathbb{E}_{XZ}[X_2(T)]^2+\mathbb{E}_{XZ}[Y_2(T)]^2\}}{2} \label{eq:chi22JJ}.
\end{align}

To isolate the decay rates $\chi_{1,1;2,2}(T)$ and $\chi_{1,2}(T)$, we rely on linear combinations of  the two-qubit expected values $\mathbb{E}_{XX}[X_1X_2(T)]$, $\mathbb{E}_{XX}[Y_1Y_2(T)]$, $\mathbb{E}_{XX}[X_1Y_2(T)]$, and $\mathbb{E}_{XX}[Y_1X_2(T)]$,
\begin{align}
\Gamma_1(T)\equiv&\,\mathbb{E}_{XX}[X_1X_2(T)]+ \mathbb{E}_{XX}[Y_1Y_2(T)]\\
=&\,e^{-\chi_{1,1;2,2}(T)+\chi_{1,2}(T)}\cos[\theta_{1}(T)-\theta_{2}(T)],\notag\\
\Gamma_2(T)\equiv&\,\mathbb{E}_{XX}[X_1X_2(T)]- \mathbb{E}_{XX}[Y_1Y_2(T)]\\
=&\,e^{-\chi_{1,1;2,2}(T)-\chi_{1,2}(T)}\cos[\theta_{1}(T)+\theta_{2}(T)],\notag\\
\Gamma_3(T)\equiv&\,\mathbb{E}_{XX}[X_1Y_2(T)]+\mathbb{E}_{XX}[Y_1X_2(T)]\\
=&\,e^{-\chi_{1,1;2,2}(T)-\chi_{1,2}(T)}\cos[\theta_{1}(T)+\theta_{2}(T)],\notag\\
\Gamma_4(T)\equiv&\,\mathbb{E}_{XX}[Y_1X_2(T)]-\mathbb{E}_{XX}[X_1Y_2(T)]\\
=&\,e^{-\chi_{1,1;2,2}(T)+\chi_{1,2}(T)}\cos[\theta_{1}(T)-\theta_{2}(T)].\notag
\end{align}
In terms of these two-qubit expected values, the decay rates can be expressed as
\begin{align}
&\chi_{1,1;2,2}(T)=-\frac{\text{ln}[\Gamma_2(T)^2+\Gamma_3(T)^2]+\text{ln}[\Gamma_1(T)^2+\Gamma_4(T)^2]}{4},\label{eq:chi1122}\\
&\chi_{1,2}(T)=\frac{-\text{ln}[\Gamma_2(T)^2+\Gamma_3(T)^2]+\text{ln}[\Gamma_1(T)^2+\Gamma_4(T)^2]}{4}. \label{eq:chi12}
\end{align}

Equations \eqref{eq:chi11JJ}-\eqref{eq:chi22JJ} and \eqref{eq:chi1122}-\eqref{eq:chi12} enable us to obtain the dynamical contributions of the spectra independent of $\Delta_1$, $\Delta_2$, and $J$. However, they also can be exploited to extract the static contributions. From the expectation values discussed above, we obtain the static detuning and static crosstalk contributions from
\begin{align}
    2T\Delta_{n} &= \tan^{-1}{\left(\frac{\mathbb{E}_{XX}[Y_{2}(T)]}{\mathbb{E}_{XX}[X_{2}(T)]}\right)} \label{eq:static-delta} \\
    2TJ &= \cos^{-1}{\left(\frac{A_{2}^{XX}(T)}{A_{2}^{ZX}(T)}\right)}.
    \label{eq:static-J}
\end{align}
We note that to accurately estimate $\Delta_{n}$ and $J$, these parameters must meet the condition $-\pi\leq 2T\Delta_{n}\leq \pi$ and $0\leq 2TJ\leq 2\pi$. If the static noise terms are very large, they cannot be accurately reconstructed with our two-point approach. We discuss this further in Appendix~\ref{app:detuning-crosstalk}, however, we note that this is an equivalent effect to phase wrapping observed in other quantum sensing applications~\cite{degen2017qs}.

\begin{figure*}
    \centering
    \includegraphics[width=\textwidth]{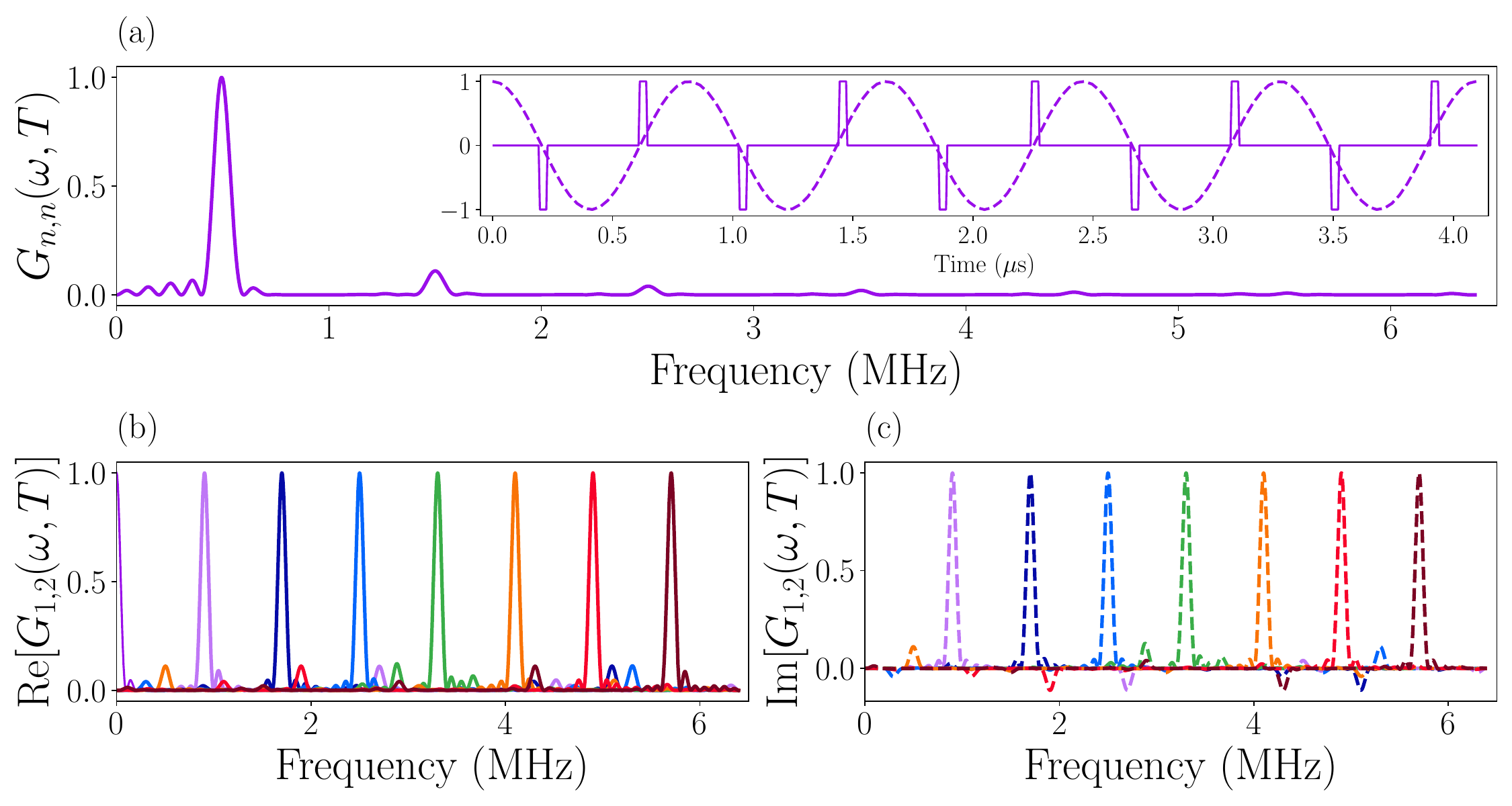}
    \caption{(a) The filter function $G_{n,n}(\omega,T)$ for a \acrshort{fttps} with $k=5$. (\textit{inset}) The cosine function (dashed) for the pulse sequence (solid) yielding the filter function in (a). The (b) real and (c) imaginary parts of the filter function $G_{0,1}(\omega,T)$ generated by the cos-cos and cos-sin switching functions, respectively. The imaginary part of the cos-cos and real part of the cos-sin filter functions are zero and not shown here.}
    \label{fig:ff}
\end{figure*}

% Control sequences sections
% Discussion on what FTTPS to use or other free evolution to ensure that noise enters the qubit dynamics
\subsection{Control Sequences and Reconstruction Protocol} \label{sec:ControlSec}
By combining expected values of Pauli observables obtained for different state preparations, we have shown that it is possible to isolate the spectra-dependent decay rates. For QNS, however, we also need a method to extract the frequency-domain characteristics of the spectra. To accomplish this, we leverage the overlap-integral structure of the decay rates along with targeted control sequences to shape the FFs. For a control sequence $i$, the decay rates implicitly defined in Eqs.~\eqref{eq:1qDecay} and \eqref{eq:2qDecay} take the general form
\begin{align}
\chi^{(i)}(T)=\sum_l C_l\int_0^\infty d\omega G_l^{(i)}(\omega,T)S_l(\omega),
\end{align}
where $C_l$ is a constant, $G_l^{(i)}(\omega,T)$ is a \acrshort{FF} generated by sequence $i$, and $S_l(\omega)$ is the corresponding spectrum. Discretizing the frequency domain overlap integral into increments 
$\Delta\omega_0\equiv[\omega_0,\omega_1],\ldots,\Delta\omega_{N}=[\omega_N,\omega_{N+1}] $ produces
\begin{align}
\chi^{(i)}(T)\approx &\,\sum_l C_l\sum_{j=0}^N\int_{\omega_j}^{\omega_{j+1}}d\omega G_l^{(i)}(\omega,T)S_l(\omega)\notag\\
\approx &\, \sum_l C_l\sum_{j=0}^NS_{l,j}\int_{\omega_j}^{\omega_{j+1}}d\omega G_l^{(i)}(\omega,T),\notag\\
= &\, \sum_l \sum_{j=0}^NS_{l,j}G_{l,j}^{(i)}(T).\label{eq:LinEq}
\end{align}
In the last line, $G_{l,j}^{(i)}(T)\equiv C_l \int_{\omega_j}^{\omega_{j+1}}d\omega G_l(\omega,T)$ and $S_{l,j}\approx S_l(\omega)$ for $\omega_j\leq\omega\leq\omega_{j+1}$, provided $S_l(\omega)$ does not vary appreciably within $\Delta\omega_{j}$. 

Observe that Eq.~\eqref{eq:LinEq}  is a linear equation that relates an experimentally measurable quantity, $\chi^{(i)}(T)$, to known FF-dependent coefficients, $G_{l,j}^{(i)}(T)$, and unknown values of the spectra within the frequency increments, $S_{l,j}$. Measuring the decay rates for a set of control sequences $i=1,\ldots,M$ generates a system of linear equations, which we can compactly write as
\begin{align}
\vec{\chi}=\mathbf{G}\vec{S}.
\end{align}
Here, $\vec{\chi}=[\chi^{(1)}(T),\ldots,\chi^{(M)}(T)]^T$ is a vector of measured decay rates, $\mathbf{G}$ is the ``reconstruction matrix" where the $i$th row contains the FF-dependent coefficients $G_{l,j}^{(i)}(T)$, and $\vec{S}$ is a vector of the $S_{l,j}$. Provided $\mathbf{G}$ is a well-conditioned matrix,  we can estimate the values of the spectra within each frequency increment by solving the system for $\vec{S}$. A necessary condition for a well-conditioned $\mathbf{G}$ is that $G_{l,j}^{(i)}(T)\neq 0$ for at least one control sequence $i$, which ensures that $S_{l,j}$ contributes to the dynamics. The conditioning is further improved when the rows of $\mathbf{G}$ are near orthogonal, which occurs when the control sequences $i=1,\ldots,M$ generate sets of \acrshort{FF}s with minimal overlap in the frequency domain \cite{norris2016qubit}. 

To generate a well-conditioned $\mathbf{G}$, we rely on control sequences composed of the cosine and sine \acrshort{fttps} \cite{PhysRevLett.131.210802}.  
The cosine FTTPS of order $k\in\{0,\ldots,2^m-1\}$ defined over a total time $T=2^{m+1}\tau_\pi$, with $\tau_\pi$ the duration of a $\pi$-pulse, consists of $2k$ equally-spaced $\pi$-pulses. The $\pi$-pulses are applied to the qubit at times $\{t_1^\text{c},...,t_{2k}^\text{c}\}$, where $t_{\ell}^\text{c}$ is the $\ell$th zero of $\text{cos}(2\pi k t/T)$ rounded to the nearest integer multiple of $\tau_\pi$.  Similarly, the sine FTTPS of order $k\in\{1,\ldots,2^m-1\}$ defined over a total time $T=2^{m+1}\tau_\pi$ consists of $2k$ equally-spaced $\pi$-pulses applied at times $\{t_1^\text{s},...,t_{2k}^\text{s}\}$, where $t_{\ell}^\text{s}$ is the $\ell$th zero of $\text{sin}(2\pi k t/T)$ rounded to the nearest integer multiple of $\tau_\pi$. The lowest order cosine and sine FTTPS are effectively CPMG and spin echo sequences, respectively, with the pulse times and intervals of free evolution discretized into integer multiples of the $\pi$-pulse duration. 

As shown in Fig.~\ref{fig:ff}, applying the cosine or sine FTTPS of order $k$ to qubit $n$ generates a $G_{n,n}(\omega,T)$ filter with a peak centered at $\omega \approx 2\pi k/T$. This ensures that $G_{n,n}(\omega,T)$ is nonzero over the frequency increment $\Delta\omega_n$ containing $2\pi k/T$. The $G_{n,n}(\omega,T)$ filters generated by FTTPS of different orders $k$, furthermore, have limited spectral overlap. For a well-conditioned $\mathbf{G}$, however, these conditions must also be satisfied by the remaining FFs -- $G_{12,12}(\omega,T)$, $\text{Re}[G_{1,2}(\omega,T)]$, and $\text{Im}[G_{1,2}(\omega,T)]$. As these are multi-qubit FFs, we identify combinations of FTTPS that when applied to qubits 1 and 2 generate peaked filters similar to $G_{n,n}(\omega,T)$. 

Consider first $G_{12,12}(\omega,T)$, which couples to the crosstalk noise spectrum and depends on the switching function $y_{12}(t)$. If we apply the $k=0$ cosine FTTPS or, equivalently, free evolution (no $\pi$-pulses) to qubit 1, then $y_1(t)=1$ and $y_{12}(t)= y_2(t)$. It follows from \eqref{eq:FFdefinition} that when we apply a cosine or sine FTTPS to qubit 2, $G_{12,12}(\omega,T)$ is equivalent to $G_{2,2}(\omega,T)$ and possesses the same favorable properties. We can alternatively apply a cosine or sine FTTPS to qubit 1 while qubit 2 undergoes free evolution, producing a $G_{12,12}(\omega,T)$ filter equivalent to $G_{1,1}(\omega,T)$.

We next consider $\text{Re}[G_{1,2}(\omega,T)]$ and $\text{Im}[G_{1,2}(\omega,T)]$. The real and imaginary components of $G_{1,2}(\omega,T)$ are determined by time-domain symmetries of the switching functions \cite{szankowski2016:2qqns,pazsilva2017:2qqns}. To illustrate this, we introduce the time-shifted switching functions $Y_{1}(t)\equiv y_{1}(t+T/2)$ and $Y_{2}(t)\equiv y_{2}(t+T/2)$, which are defined on $t\in[-T/2,T/2]$. If $f(t)$ is a time-dependent function, we denote its odd and even components by $\mathcal{E}f(t)\equiv \left[f(t)+f(-t)\right]/2$ and $\mathcal{O}f(t) \equiv \left[f(t)-f(-t)\right]/2$, respectively. It follows from \eqref{eq:FFdefinition} that $G_{1,2}(\omega,T)$ is given by 
\begin{align}\label{eq:G12}
    G_{1,2}(\omega,T)&= \left[\int_{-T/2}^{T/2} dt \,Y_{1}(t)e^{i\omega t}\right] \left[\int_{-T/2}^{T/2} dt \,Y_{2}(t)e^{-i\omega t}\right]\nonumber\\
    &= \left\{ \int_{-T/2}^{T/2} dt \left[\mathcal{E}Y_{1}(t)\cos{\omega t}+i\mathcal{O}Y_{1}(t)\sin{\omega t}\right]\right\} \nonumber\\
    &\times\left\{ \int_{-T/2}^{T/2} dt \left[\mathcal{E}Y_{2}(t)\cos{\omega t}-i\mathcal{O}Y_{2}(t)\sin{\omega t}\right]\right\}.
\end{align}
Note that $G_{1,2}(\omega,T)$ is real if $Y_{1}(t)$ and $Y_{2}(t)$ are both odd or both even in $t$. On the other hand, $G_{1,2}(\omega,T)$ is imaginary if $Y_{1}(t)$ and $Y_{2}(t)$ have opposite parity, i.e., they are odd and even or even and odd, respectively.

Leveraging the symmetries of the cosine and sine FTTPS switching functions enables us to design real and imaginary components of $G_{1,2}(\omega,T)$ that produce a well-conditioned $\mathbf{G}$. 
When applied to qubit $n$, the cosine FTTPS results in a time-shifted switching function $Y_{n}(t)$ that is largely even, while the sine FTTPS yields a time-shifted switching function that is largely odd. By applying identical cosine or sine FTTPS of order $k$ to both qubits, we produce an FF such that $\text{Im}[G_{1,2}(\omega,T)]\approx 0$ and $\text{Re}[G_{1,2}(\omega,T)]= G_{1,1}(\omega,T)=G_{2,2}(\omega,T)$. To generate an FF with a large imaginary component, we apply  a cosine FTTPS of order $k$ to qubit 1 (2) and a sine FTTPS of order $k$ to qubit 2 (1). In this case, $\text{Re}[G_{1,2}(\omega,T)]\approx 0$ and $\text{Im}[G_{1,2}(\omega,T)]$ has a peak centered at $2\pi k/T$. For different $k$, $\text{Im}[G_{1,2}(\omega,T)]$ has minimal spectral overlap, similar to $G_{1,1}(\omega,T)$ and $G_{2,2}(\omega,T)$.  Examples of the real and imaginary FFs for different $k$ values are shown in Figure~\ref{fig:ff}. We further expand on the inversion procedure in Appendix~\ref{linarinv}.

\subsection{Summary of FTTPS-Based Protocol}
\label{subsec:sum-protocol}

Here, we summarize the protocol used to learn both static and dynamic components of the noise. The sequences to obtain the required noise coefficients are the following:
\begin{enumerate}
    \item Cosine \acrshort{fttps} of order $k$ applied to qubit 1, free evolution on qubit 2 for $k\in\{0,...,K-1\}$.
    \item Free evolution on qubit 1, cosine \acrshort{fttps} of order $k$ applied to qubit 2 for $k\in\{0,...,K-1\}$.
    \item Cosine \acrshort{fttps} of order $k$ applied to both qubit 1 and qubit 2 for $k\in\{0,...,K-1\}$.
    \item Cosine \acrshort{fttps} of order $k$ applied to qubit 1, sine \acrshort{fttps} of order $k$ applied to qubit 2 for $k\in\{1,...,K-1\}$.
\end{enumerate}
The qubits are prepared in either $|+X\rangle$ or $|+Z\rangle$ and measured along $\{X,Y\}$. Information about the different state preparations and measurements are further outlined in Appendix~\ref{sequences}.

\section{Simulations}\label{sec:Simulations}
In this section, we evaluate the efficacy of our approach via numerical simulation. \acrshort{SchWARMA} models are used to generate spatiotemporally correlated dephasing noise for our comparisons. The inversion technique described in Sec.~\ref{sec:Protocol} is used to estimate self- and cross-spectra for the two-qubit model, however, we note that \acrshort{SchWARMA} models may also be used for parametric estimation of noise spectra~\cite{schultz2020schwarma}. The estimated statistical quantities---including means and noise spectra---are compared against theoretical values. This analysis provides a baseline comparison to explore the efficacy of the protocol prior to experimental demonstration.

\begin{figure*}
    \centering
    \includegraphics[width=\textwidth]{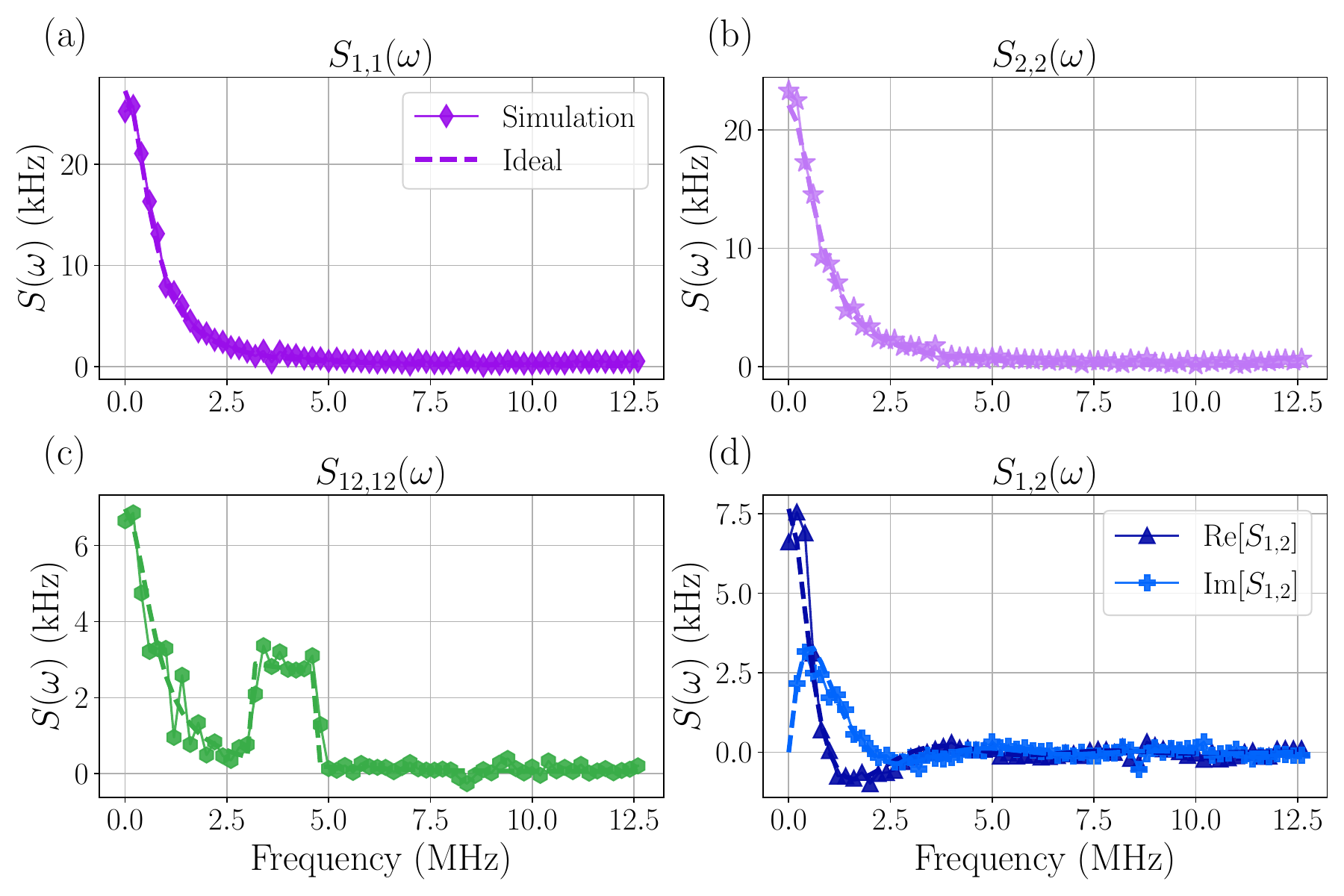}
    \caption{Simulation of the noise spectral estimates for correlated noise with a delay of $T/20$ and correlated crosstalk. The estimated spectra (marker) are in good agreement with the engineered spectra. (a-b) The single-qubit noise power spectrum for $q_{0}$ and $q_{1}$, respectively. (c) The crosstalk noise spectrum with Lorentzian spectrum and bandpass feature. (d) The real and imaginary components of the cross-spectrum.}
    \label{fig:sim-spec}
\end{figure*}

\subsection{Simulation Specifications}
\label{subsec:sim-specs}
In our simulations, each noise process in Eq.~\eqref{eq:H-int} is generated by an Ornstein-Uhlenbeck (OU) process. The OU model, a widely used model for noise in quantum systems~\cite{hanggi1985bistability:OU,wilkinson2010perturbation:OU}, is described by the stochastic process
\begin{equation}
\eta(t + \delta t) = \eta(t) + \theta\left[\mu - \eta(t)\right] \delta t + \sigma \sqrt{2\theta}\, \xi(t)
\end{equation}
where $\mu$ is the mean, $\sigma$ is the stationary standard deviation, and $\theta$ is effectively related to the correlation time of the noise $\tau_c\sim1/\theta$. Here, we consider a deterministic start with $\eta(0)=0$, while $\xi(t)$ is generated from a normal distribution $\mathcal{N}(0, \delta t)$. $\delta t$ denotes the time resolution for the control. The Lorentzian
\begin{equation}
S_L(\omega) = \frac{2\theta \sigma^2}{\theta^2 + \omega^2}
\end{equation}
describes the PSD of the OU process.

The local dephasing processes for each qubit are defined as linear combinations of three OU processes. Specifically, $\beta_1(t)=\eta_1(t) + \eta_3(t)$ and $\beta_2(t+\tau) = \eta_2(t) + \eta_3(t+\tau)$, where $\eta_j(t)$, $j=1,2,3$ are independent OU processes. In this way, we can generate spatiotemporally correlated noise between the local dephasing noise processes. The resulting self-spectra are composed of two Lorentzians, while the cross-spectra is dictated by the specifications of $\eta_3(t)$ alone. Moreover, the additional delay of $\tau$ allows for temporal shifts in the cross-spectrum, and subsequently, the emergence of both real and imaginary components in the cross-spectrum. 

The crosstalk spectrum consists of a Lorentzian spectrum together with an additional bandpass feature. The latter is modeled by the PSD
\begin{equation}
S_{BP}(\omega) = A_0 \left[ \Theta(\omega - \omega_{\ell}) - \Theta(\omega - \omega_{h})\right],
\end{equation}
where the amplitude $A_0$ sets the feature height, and the low and high-frequency cutoffs, $\omega_\ell$ and $\omega_h$, respectively, determine its width.  $\Theta(\omega)$ is the Heaviside function, which takes the value of unity for $\omega>0$ and zero otherwise. The complete crosstalk spectrum is given by the sum of the Lorentzian background and this bandpass contribution. We introduce this feature to highlight the QNS protocol's ability to estimate non-trivial PSDs. Without loss of generality, we select the crosstalk spectra for the inclusion of the bandpass feature.

Upon specifying the parameters for each noise process, we convert the noise models to SchWARMA models. This enables fast simulation of the OU processes with greater numerical stability. We provide further information on the simulation environment in Appendix~\ref{app:schwarma}.

\subsection{Spectrum Reconstruction Analysis}
In Fig.~\ref{fig:sim-spec}, we compare the ideal noise spectra---defined according to the following parameters---and the noise spectra estimated by our QNS protocol.  The local dephasing noise for each qubit is specified by $\eta_j(t)$ as described in Sec.~\ref{subsec:sim-specs}. The parameters for each OU noise process are given by $\sigma_1 T=1.05$ and $\theta_1 T=22.5$, $\sigma_2 T=0.99$ and $\theta_2 T=27.5$, and $\sigma_3 T=0.62$ and $\theta_3 T=20$, where the total time $T=5\mu s$. A delay between the local dephasing noise processes of $\tau=T/20$ is included as well. The qubit detuning is set to $\Delta_\mu T=0.2\pi$ for $\mu=1,2$. Crosstalk noise is characterized by a Lorentzian with parameters $\sigma_{12} T=0.65$ and $\theta_{12}T=24$, with static error $JT=0.6\pi$. The bandpass feature possesses an amplitude of $A_0 T=0.0125$, with cutoff frequencies $\omega_\ell T=100$ and $\omega_h T=150$. The system is subject to the \acrshort{fttps} with $m=6$, which implies a gate time of $\delta t \approx 40$ ns. $1000$ noise realizations are used to estimate the noise-averaged expectation value for each state and observable required by the protocol.

As can be seen from the comparison, the estimated spectra agree well with the expected values. We quantify this agreement through the mean absolute error (MAE): $\Gamma = \frac{1}{n}\sum_{i=1}^{n}|S_{\rm est}(\omega_i)-S_{\rm eng}(\omega_i)|$, which is determined by the estimated spectrum $S_{\rm est}(\omega)$ and engineered spectrum $S_{\rm eng}(\omega)$. The engineered noise spectrum refers to the noise used in the simulation, while in Sec.~\ref{sec:Experiments} it will refer to noise injected into the experimental system. Here, the MAE for the dephasing self-spectra is $268$ Hz for $q_1$ and $330$ Hz for $q_2$, and relative to the range of the engineered noise this is an error of approximately $0.9\%$ and $1.3\%$. The error in the cross-spectra is $254$ Hz and $86$ Hz for real and imaginary, respectively. Relative to the noise range this constitutes a deviation of $3.4\%$ for the real spectrum and $0.9\%$ for the imaginary spectrum. The crosstalk MAE is $111$ Hz, corresponding to a deviation of $2.8\%$. This is slightly larger due to the presence of the bandpass feature.

In addition to the spectral reconstructions, we use our simulations to estimate the static noise terms using Eqs.~\eqref{eq:static-delta} and \eqref{eq:static-J}. Here, we find a relative error of 0.058 and 0.056 for the detuning rates and 0.033 for the static crosstalk. 

\begin{figure}
    \centering
    \includegraphics[width=\columnwidth]{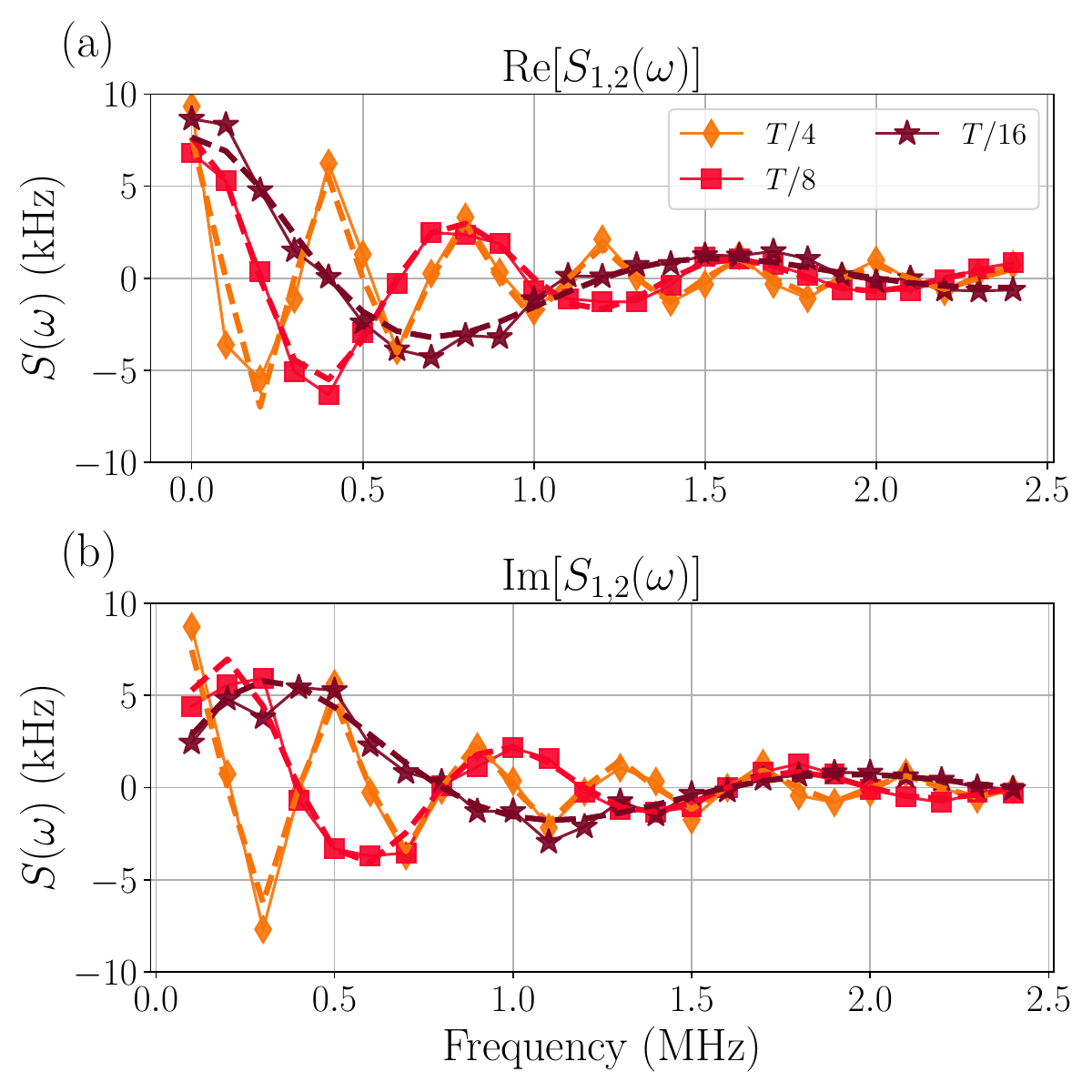}
    \caption{The (a) real and (b) imaginary cross-spectra at four delay times $\{T/4, T/8, T/16\}$, where $T$ is the total time of the \acrshort{fttps}. The oscillation frequency increases with longer delay between the two noise processes on the two qubits. The dashed lines show the engineered spectra.}
    \label{fig:sim-delay}
\end{figure}

In Fig.~\ref{fig:sim-delay}, we include an additional analysis for a nonzero delay in the cross-correlation function. We choose similar parameters to the simulations above, however, we consider varying $\tau$ values and choose $T=10\mu$s. The estimated real and imaginary components of the cross-correlation PSD are shown to strongly overlap with the engineered spectra. Computing the MAE, we find an error of $287$ Hz and $246$ Hz for both the real and imaginary components, respectively, for $\tau=T/16$. Larger delays result in similar error magnitudes. For $\tau=T/8$, the real and imaginary spectra differ from the engineered by an error of $240$ Hz and $246$ Hz, while for $\tau=T/4$, the error magnitudes are $327$ Hz and $243$ Hz. Relative to the engineered spectrum, these errors are $\approx2\%$ of the range. In all cases, we find that the estimated spectra agree well with the expected.

\section{Experiments}\label{sec:Experiments}
We experimentally demonstrate two-qubit noise spectroscopy by measuring an engineered noise environment on a pair of neighboring qubits. The expectation values are used to reconstruct the single-qubit noise spectra $S_{n,n}(\omega)$, cross-spectra $S_{n,n'}(\omega)$, and crosstalk noise spectrum $S_{12,12}(\omega)$. Reconstructed spectra are validated by comparison to our model which includes the engineered noise and contributions from the independently measured, native background. Overall, we find good agreement between the measured and modeled noise spectra.

\subsection{Device}

The experiments were performed on a superconducting qubit processor consisting of six fixed-frequency transmons arranged in a ring with nearest-neighbor capacitive coupling. Each qubit has an independent control line and readout resonator. The readout resonators are coupled to a single readout bus and measurement tones are multiplexed. Two-qubit QNS was implemented on a pair of neighboring qubits $(q_1, q_2)$ with frequencies $\omega_{q_1}/2\pi=5.84$ GHz and $\omega_{q_2}/2\pi=6.07$ GHz, and coherence times $T_{1}^{1q_1}=28$ $\mu$s ($T_{1}^{q_2}=34$ $\mu$s) and $T_{2}^{q_1}=37$ $\mu$s ($T_{2}^{q_2}=34$). Gaussian DRAG-corrected $\pi-$ and $\pi/2$-pulses, of duration $t_{g}=32$ns, were calibrated for both qubits.

% Figures on delay vs no delay
\begin{figure*}
    \centering
    \includegraphics[width=\textwidth]{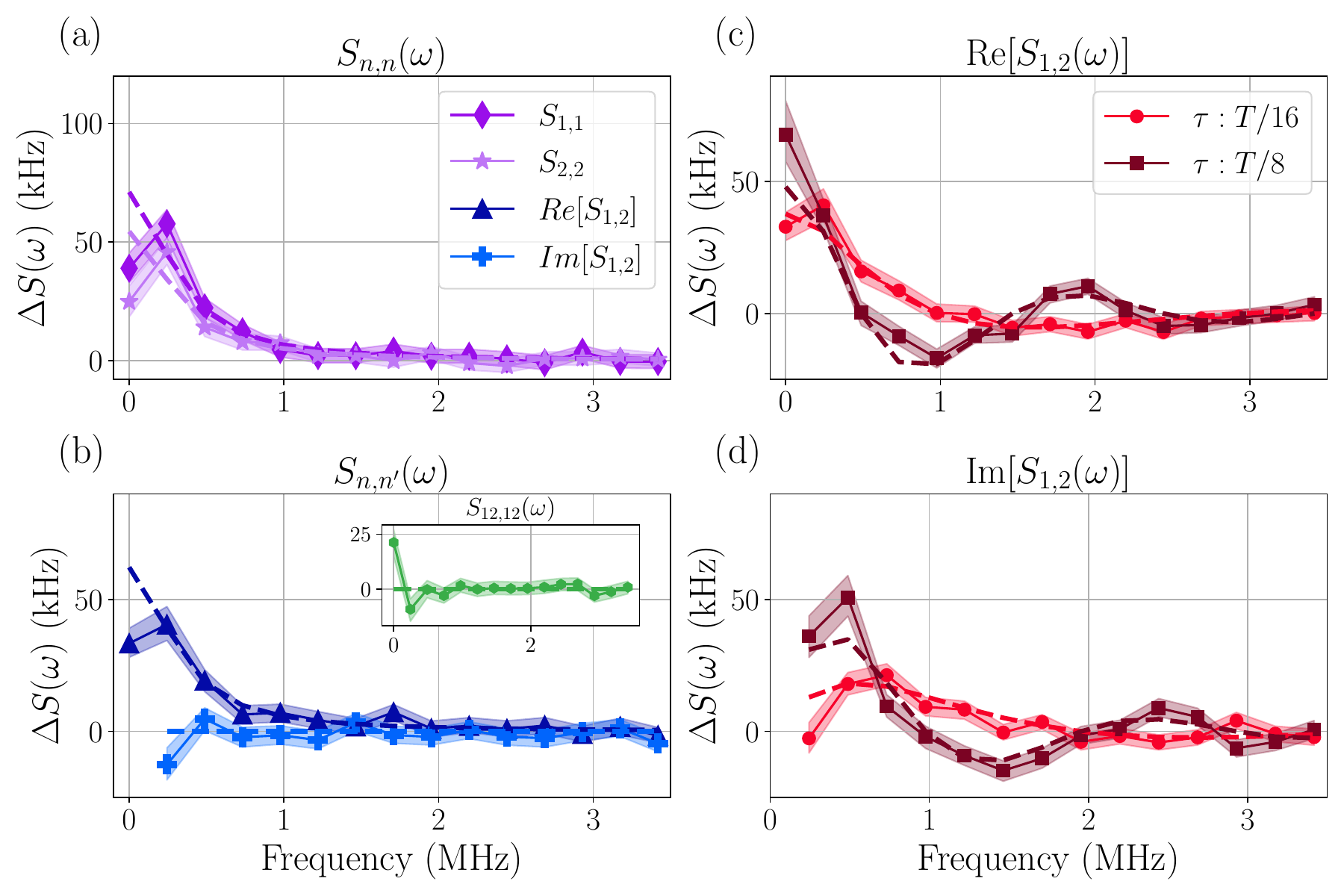}
    \caption{ Reconstructed two-qubit noise spectra for engineered Lorentzian noise, averaging over 50 noise trajectories, with the native-noise baseline subtracted. The dashed lines are the engineered noise spectra, solid lines with markers are the reconstructed spectra, and shaded regions are the CI determined with bootstrapping. Reconstructed (a) single-qubit spectra and (b) cross-spectra for Lorentzian noise with no time delay between the two qubits' noise trajectories. The crosstalk spectrum (inset) shows noise at lower frequencies. The (c) real and (d) imaginary cross-spectra for different delays between the two noise trajectories injected on both qubits. }
    \label{fig:correlated_noise}
\end{figure*}

To implement the two-qubit QNS protocol, we apply simultaneous $\pi$-pulses during the cosine \acrshort{fttps} on both qubits and non-simultaneous $\pi$-pulses  when playing cosine \acrshort{fttps} on qubit 1 and sine \acrshort{fttps} on qubit 2. For sequences involving simultaneous control of both qubits, adverse classical crosstalk effects were mitigated by using simultaneously calibrate $\pi$-pulses. 

For these experiments, the total time of the FTTPS was kept below $T_{1}/10$ such that $T_{1}$ would not contribute appreciably to the dynamics. In order to stay in the fast-pulse limit, we implemented only those FTTPS for which the delay between pulses was $\geq 3t_g$.

% Protocol validation section
\subsection{Protocol Validation}
First, we experimentally validated our protocol by using it to estimate engineered spatiotemporally correlated dephasing noise.  The correlated phases were generated with SchWARMA and implemented in the experiment using the gate-based noise injection technique introduced in Ref.~\cite{murphy2021universal}. By averaging over many noise trajectories, the desired dephasing noise spectrum can be observed. 

For these experiments, we generated $50$ unique noise trajectories with $\sigma T=1.36$ and $\theta T=8.2$, and $100$ shots were collected for each trajectory.  Each noise trajectory was applied to all sequences with different scaling factors depending on the qubits. We chose  $\beta_{1}(t)= 0.8\eta(t)$ for the first qubit and $\beta_{2}(t+\tau)= 0.8\eta(t+\tau)$ for the second qubit, where $\eta(t)$ is a common noise process among the two qubits, and $\tau$ is a relative delay.

% Injecting engineered noise
% \subsection{Injecting Engineered Noise}

The total evolution time of the sequences was $T=4.096$ $\mu$s and $m=6$, corresponding to $2^{m}=64$ sequences containing $2^{(m+1)}=128$ pulses. The Nyquist frequency as set by the gate time is $\omega_{\rm Nyq}/2\pi\equiv1/2t_g\approx15.6$ MHz with a bandwidth of $\delta\omega/2\pi\approx0.244$ MHz, determined by the total time.  The QNS procedure produces estimates of the spectra across the frequency band $\omega\in [0, 2\pi K/T]$, where $K$ is the maximum FTTPS order. As detailed in Sec. \ref{sec:ControlSec}, the spectral estimates are solutions to a linear system dependent on the measured observables and control sequences applied to the qubits. The linear system generated by our particular choice of observables and sequences is given in Appendix~\ref{linarinv}.

The reconstructed spectra for the engineered noise without a delay (i.e., $\tau=0$) is shown in Fig.~\ref{fig:correlated_noise} (a)-(b). The average spectrum estimate (solid line) and confidence intervals (CIs) (shaded) are plotted with the modeled spectrum (dotted lines). We isolate the engineered noise spectrum by independently measuring the native noise spectrum $S_{\rm nat}(\omega)$ and subtracting it from the reconstructions that include the engineered noise $S_{\rm inj+nat}(\omega)$. This results in the effective estimate of the engineered noise spectrum $\Delta S(\omega) = S_{\rm inj+nat}(\omega) - S_{\rm nat}(\omega)$. Measurement error mitigation (MEM) based on iterative Bayesian unfolding~\cite{pokharel2024scalable} was applied to the averages over single-shot outcomes and the $95\%$ CIs were determined via bootstrapping with replacement from resamples of the data~\cite{stine1989introduction:95ci}.

The reconstructed spectra are within error margins of the expected spectrum for most frequencies. The first point in the spectrum corresponds to a Ramsey sequence with no echo pulses. Drifts in the qubit frequency have a larger impact on the noise spectrum at DC, leading to larger deviations from the modeled noise spectrum. However, overall, the spectra agree quite well with the expected noise profiles shown in Fig.~\ref{fig:correlated_noise} (a)-(b). The MAE for the single-qubit spectrum is $4.1$ kHz ($6\%$ of the range) and $3.9$ kHz ($7\%$ of the range) for $q_{1}$ and $q_{2}$, respectively. The real part of the cross-spectrum has an MAE value of $3.3$ kHz ($0.5\%$ of the range), while the imaginary cross-spectrum MAE is $3$ kHz. Finally, the crosstalk spectrum, inset of Fig.~\ref{fig:correlated_noise} (b) has an MAE of $3.1$ kHz. The range is zero for the engineered imaginary cross-spectrum and crosstalk spectrum, but the error magnitudes are similar to the other estimated spectra.

As discussed in Section~\ref{sec:Simulations}, our protocol estimates the single-qubit and crosstalk static noise terms as well. For these experiments, we sought to measure the native static noise terms, and thus all engineered noise processes were zero-mean. For the experiments shown in Fig.~\ref{fig:correlated_noise}(a), we report $\Delta_{1}=-29$ kHz and $\Delta_{2}=-11$ kHz. We expect the detuning to be on the order of a few kHz and while the estimated detuning is larger, we note that it is averaged over a two hour period. Our device has a static crosstalk $JT \gg 2\pi$, which does not meet the crosstalk condition, and so the crosstalk estimate falls into the phase wrapping regime. The crosstalk value from \acrshort{QNS} is $J=28$ kHz, while the value estimated using the joint amplification of $ZZ$ method~\cite{takita2017experimental:jazz} is $380$ kHz.

One of the highlights of the two-qubit \acrshort{QNS} protocol is the time-asymmetric correlations between the two qubits which is quantified by the imaginary cross-spectrum. We introduce a delay by shifting the phase trajectory of the second qubit relative to those of the first by $T/8$ and $T/16$. A larger delay will result in a larger component of the imaginary noise spectrum. Fig.~\ref{fig:correlated_noise} (c)-(d) shows the cross-spectrum estimates with two different delay values. Parameters for the Lorentzian noise are $\sigma_{\tau/8}T=1.3$ and $\theta_{\tau/8}T=20.5$, and $\sigma_{\tau/16}T=1.4$ and $\theta_{\tau/16}T=18.4$ to accentuate the difference in delay time.

While we observe good agreement between the engineered and reconstructed spectra, several factors contribute to discrepancies. Noise injection experiments for a specific noise spectrum are executed over a two-hour period with interleaved experiments that probe the native noise.  While the native noise measurements are used to subtract the background from the reconstruction of the engineered spectrum, they may not accurately capture the native spectrum since they occur every 12 minutes. Fluctuations of $T_{1}$ and $T_{2}$ have been shown to occur natively within this timescale in superconducting devices~\cite{burnett2019:fnoise}, leading to discrepancies between the reconstructed and expected values. Furthermore, the native noise spectrum is not flat due to single-qubit dephasing noise from two-level fluctuators or thermal photons~\cite{yan2018distinguishing}, and in some cases---as shown in Fig.~\ref{fig:memvsraw}---is comparable to the amplitude of the engineered noise. This effectively decreases the signal-to-noise ratio of the engineered noise and degrades MAE.

% Native noise
\subsection{Native Noise}

Here, we examine the native noise spectra to assess the presence of spatiotemporal correlations. These experiments were interleaved between the injected noise measurements and averaged over ~$26,000$ shots. For our pair of qubits, the single-qubit noise spectra are generally flat, though $q_{1}$ has a higher native noise spectrum than $q_{2}$, as shown in Fig.~\ref{fig:memvsraw}(a) and (b), respectively. This could be attributed to $T_{1}$ and $T_{2}$. The single-qubit noise is the dominant noise term, however, there is a small contribution from the cross-spectra; see panel (d). The imaginary cross-spectrum is on the order of the crosstalk noise and is about $10-20\%$ of the self-spectra.

\subsection{Impact of Measurement Errors}
\begin{figure}
    \centering
    \includegraphics[width=\columnwidth]{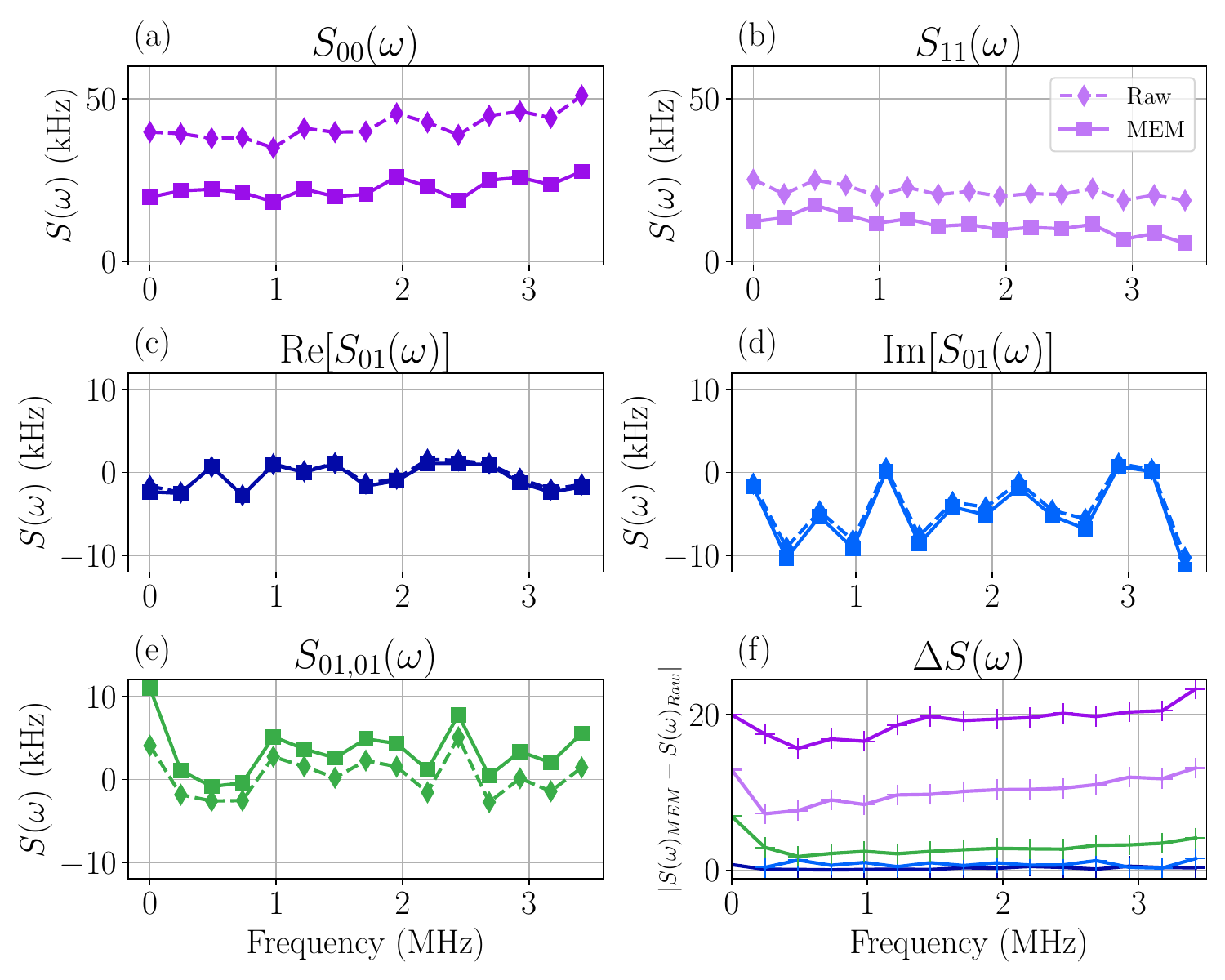}
    \caption{Difference in the reconstructed noise spectra using the measurement error mitigated counts~\cite{pokharel2024scalable} and raw counts for (a) $q_1$ and (b) $q_2$ self-spectrum, (c) real- and (d) imaginary cross-spectra, and (e) crosstalk spectrum. The different in the reconstructed noise spectra is shown in (f).}
    \label{fig:memvsraw}
\end{figure}

MEM is important for qubit systems where state prep and measurement (SPAM) errors can be large. Certain characterization techniques are robust to SPAM error, such as randomized benchmarking~\cite{emerson2005scalable} and gate set tomography~\cite{nielsen2021gate}, and thus, do not need MEM.  Recent \acrshort{QNS} studies have sought to realize SPAM robustness as well~\cite{khan2024spam}. Here, we investigate the impact of measurement error on the two-qubit \acrshort{QNS} protocol. We compare the reconstructed native noise spectra with and without MEM, as shown in Fig.~\ref{fig:memvsraw}. We find that the single-qubit and crosstalk spectra are impacted the most by MEM, where misclassification of shots changes the expectation value of the observables; see Fig.~\ref{fig:memvsraw}(a), (b), and (e). We attribute these errors to measurement infidelity of $3-8\%$ and classical crosstalk from the device design.

% Figure on raw vs mem native noise 

\subsection{Comparison to Frequency Comb Approach}
Frequency comb methods~\cite{pazsilva2017:2qqns,szankowski2016:2qqns} can suffer from poorly conditioned reconstruction matrices and may not detect narrowband noise spectra that fall within the fundamental harmonics of the sequence. Such features may heavily influence noise tailoring in control protocols~\cite{oda2023optimally}.  Here, we highlight the advantage of \acrshort{fttps} over comb-based methods in the detection of narrowband features in the noise.

\begin{figure}
    \centering
    \includegraphics[width=\columnwidth]{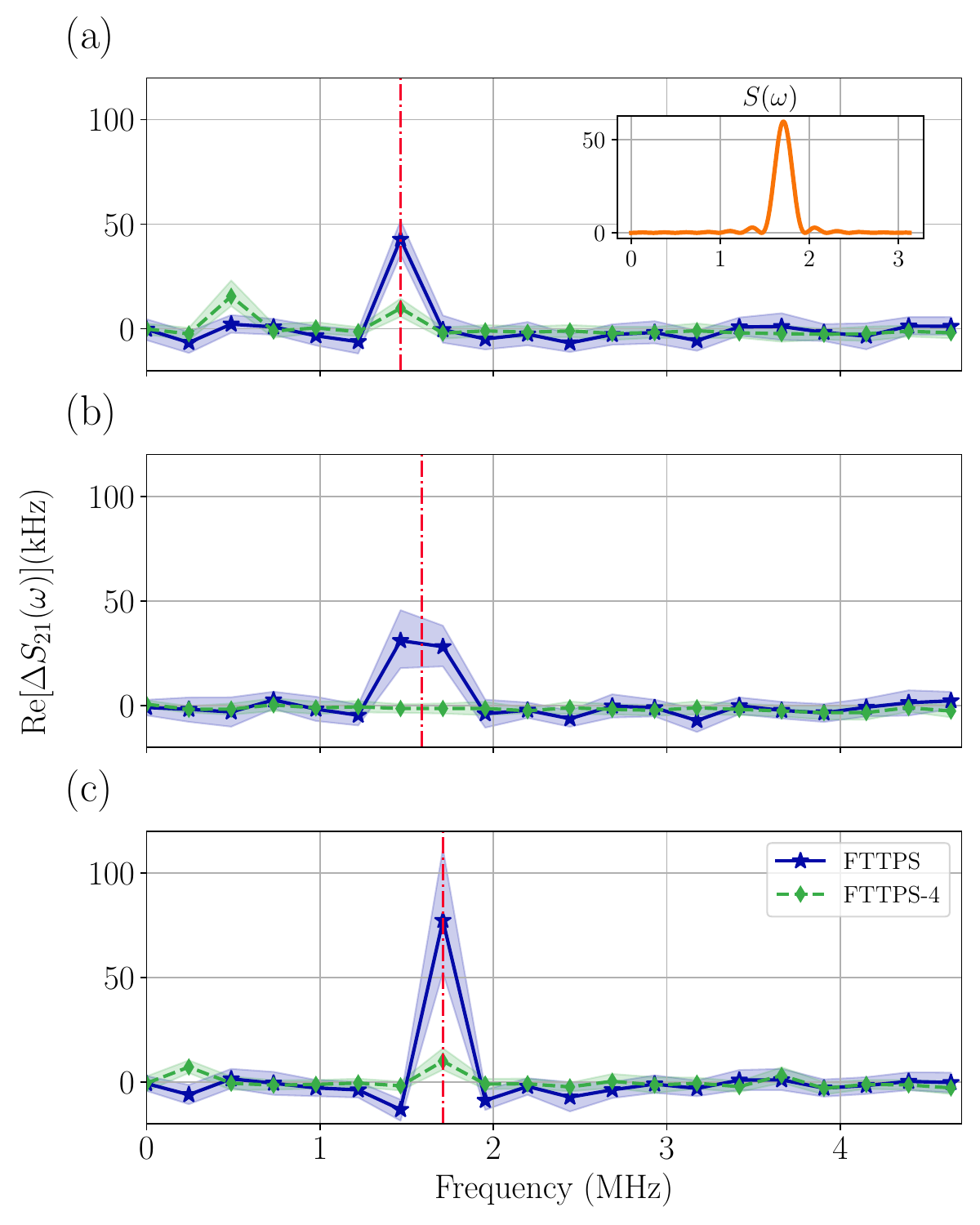}
    \caption{The real component of the cross-spectrum for the \acrshort{fttps} two-qubit \acrshort{QNS} method and \acrshort{fttps}-$M$, where $M=4$ the number of repetitions, for narrow-band noise at three different frequencies $\omega/2\pi = {6/T, 6.5/T, 7/T}$, where $T$ is the total time of the \acrshort{fttps}. The filter functions for both methods are centered around $2\pi k/T$, however, the frequency comb \acrshort{FF} are sharply peaked with minimal overlap between neighboring filters. Our approach can reconstruct sharply peaked noise that falls between the \acrshort{FF} because of the overlap between \acrshort{FF}.}
    \label{fig:fcvsfttps}
\end{figure}

In this case, we examine engineered narrowband noise peaked at $\omega/2\pi=\alpha/T$, where $\alpha$ dictates the center frequency. We generated 50 noise trajectories for $\alpha={6,6.5,7}$ and collected 100 shots per noise trajectory. The noise spectrum of the injected noise for one frequency, $\alpha=6$, is given by the inset in Fig.~\ref{fig:fcvsfttps}. Since the cosine- and sine-\acrshort{fttps} meet the conditions necessary to measure the real and imaginary components of the cross-spectra, we use these pulse sequences and repeat them $M=4$ times to implement the frequency comb protocol. In an ideal experiment, the number of repetitions would be much larger than 4 to achieve a narrow comb, however our choice of repetitions was limited by $T_{1}$ of the qubits.

The reconstructed noise cross-spectra for our approach, FTTPS (solid), and the frequency comb method, FTTPS-4 (dashed), are plotted in Fig.~\ref{fig:fcvsfttps}. When the frequency of the injected noise is an integer multiple of $2\pi/T$, the filter function of the frequency comb overlaps with the noise. However, at non-integers, the frequency comb method cannot faithfully reconstruct the noise spectrum.

We find that the MAE for $\alpha=6$ is $6.6$ kHz and $7.9$ kHz for FTTPS and FTTPS-4, respectively. The error is $10\%$ and $13\%$ of the range, respectively. The largest error occurs $\alpha=6.5$---a non-integer multiple of $2\pi/T$---with an error of $5.5$ kHz ($10\%$ of range) for FTTPS and $7.9$ kHz ($15\%$ of range) for FTTPS-4. The MAE between FTTPS and FTTPS-4 at $\alpha=7$ is $6.7$ kHz  ($11\%$ of range) for FTTPS and $7.0$ kHz ($12\%$ of range) for FTTPS-4. These large discrepancies for FTTPS-4 come from bad estimates near the noise peak. The noise estimates from FTTPS are in excellent agreement with the predicted values.

\section{Conclusion}\label{sec:Conclusion}
In this paper, we address the challenge of enabling fast multi-qubit correlated noise characterization by proposing and experimentally validating a novel two-qubit quantum noise spectroscopy protocol. We successfully reconstruct an engineered noise spectrum, including engineered time asymmetry in the cross-spectra. This enables the study of spatiotemporally correlated noise in quantum devices and extends investigations into temporal correlations of quantum crosstalk.

Our approach uses \acrshort{fttps}, a set of classical signal processing motivated pulse sequences that sweep the filter function of the control. The advantage of using \acrshort{fttps} is that they rely on standard single-qubit rotations, reducing the calibration overhead that other \acrshort{QNS} techniques require~\cite{lupke2020experiment:2qqns}. Another advantage is the probing bandwidth of the FTTPS filter functions. Compared to the frequency comb approach~\cite{szankowski2016:2qqns}, FTTPS afford greater resolution in the probing frequencies. As we demonstrated experimentally, this allows us to measure and reconstruct sharp features in the noise spectrum that may be undetected by the frequency comb technique. 

\acrshort{SchWARMA} models were used to generate and inject spatiotemporally correlated noise to validate the protocol. We show in simulation and experiments that the reconstructed spectrum is in good agreement with the ideal noise spectrum. In addition, the protocol was employed to estimate the single-qubit and crosstalk static noise terms from simulation. Experimentally, while the single-qubit noise satisfied the phase wrapping condition, the crosstalk estimation failed due to violations of this condition. We anticipate this approach could be better suited for systems with lower quantum crosstalk strengths, e.g., tunable coupler architectures~\cite{yan2018tunable:tc,stehlik2021tunable:tc}. 

Future directions for multi-qubit \acrshort{QNS} include utilizing the protocol to study correlated noise phenomena in superconducting qubits, such as quasiparticles and two-level systems. Another possible avenue is to leverage the QNS protocol in the development of multi-qubit non-Markovian noise models. Finally, model-based approaches to noise spectroscopy have shown improved spectral estimation when compared to inversion approaches~\cite{schultz2024model}. A next step would be to combine this protocol introduced here with ARMA model parametric estimation.

\section{Acknowledgements}
The authors would like to thank Andrew Murphy, Timothy Sweeney, and Christopher Watson for useful discussions. Devices were fabricated and provided by the Superconducting Qubits at Lincoln Laboratory (SQUILL) Foundry at MIT Lincoln Laboratory, with funding from the Laboratory for Physical Sciences (LPS) Qubit Collaboratory.

\appendix

\section{Pulse Width Errors}\label{app:pulsewidth_errors}
\begin{figure*}[t]
    \centering
    \includegraphics[scale=0.48]{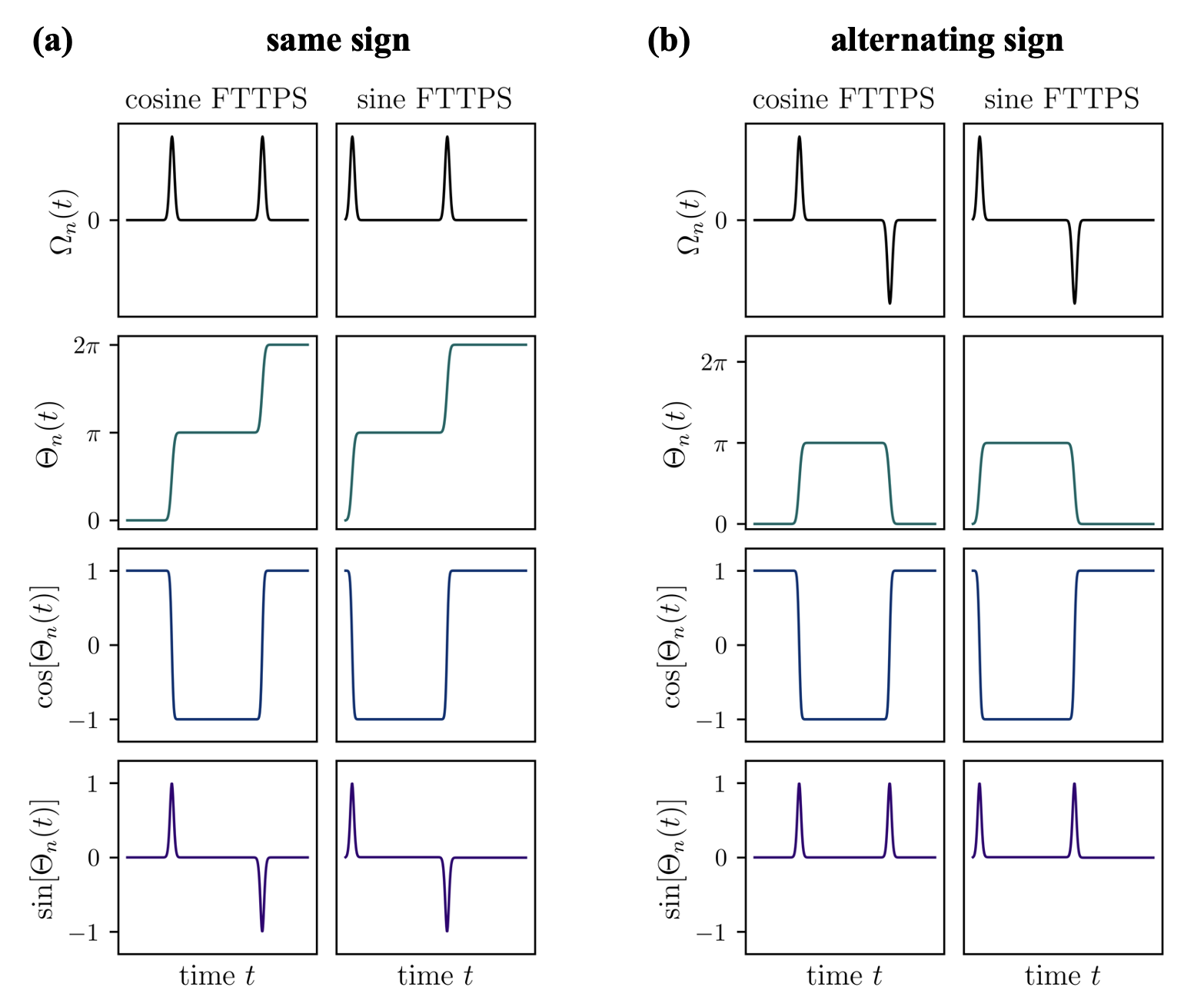}
    \vspace*{-2mm}
    \caption{{The effect of pulse waveforms on pulse width errors. Two types of waveforms are shown: (a) ``same sign", where the waveform generates consistent $\pi$-rotations along $X_n$; and (b) ``alternating sign'', where the waveform generates rotations along $X_n$ that alternate between $+\pi$ and $-\pi$. }
    }
    \label{pulsewidths}
\end{figure*}

The QNS protocol in the main text is derived under the assumption that $\pi$-pulses applied to the qubits occur instantaneously in time, ensuring the qubit dynamics are purely dephasing. In reality, however, pulses have a finite time duration, which contributes off-axis terms to the dynamics that result in errors. To illustrate this, we again transform the rotating-frame Hamiltonian $H(t)=H_\text{int}(t)+H_\text{ctrl}(t)$ into the toggling frame or interaction picture associated with the control Hamiltonian $H_\text{ctrl}(t)$. As in the main text, we consider the case where the phases of the microwave drives on qubit 1 and qubit 2 are set to zero, so that $H_\text{ctrl}(t)=\sum_{n=1}^2\Omega_n(t)X_n/2$. Accounting for the width of the pulses, the toggling frame Hamiltonian is given by
\begin{widetext}
\begin{align}
\tilde{H}(t)=&\,[\beta_1(t)+\Delta_1]\sin[\theta_1(t)]Y_1+[\beta_1(t)+\Delta_1]\cos[\theta_1(t)]Z_1\\
&+[\beta_2(t)+\Delta_2]\sin[\theta_2(t)]Y_2+[\beta_2(t)+\Delta_2]\cos[\theta_2(t)]Z_2\notag\\
&+[\beta_{12}(t)+J]\sin[\theta_1(t)]\sin[\theta_2(t)]Y_1Y_2+[\beta_{12}(t)+J]\cos[\theta_1(t)]\cos[\theta_2(t)]Z_1Z_2\notag\\
&+[\beta_{12}(t)+J]\cos[\theta_1(t)]\sin[\theta_2(t)]Z_1Y_2+[\beta_{12}(t)+J]\sin[\theta_1(t)]\cos[\theta_2(t)]Y_1Z_2,\notag
\end{align}
\end{widetext}
where the rotation angle generated by the control on qubit $n$ is $\theta_n(t)=\int_0^tds\,\Omega_n(s)$. Note that we recover the toggling-frame Hamiltonian in \eqref{eq:Htog} in the instantaneous pulse limit, where $\Omega_n(t)=\sum_{\ell=1}^{2k}\pi\delta(t-t_\ell)$ so that $\theta(t)$ is always an integer multiple of $\pi$, $\cos[\theta_n(t)]=y_n(t)\in\{+1,-1\}$, and $\sin[\theta_n(t)]=0$. 

For pulses with finite time duration,  $\sin[\theta_n(t)]$ can be nonzero. The resulting off-axis terms  along $Y_1$, $Y_2$, $Y_1Y_2$, $Z_1Y_2$ and $Y_1Z_2$ contribute error and imply that $\tilde{H}(t)$ is no longer purely dephasing, even when the pulses are perfect. Figure~\ref{pulsewidths} shows the contribution of the off-axis terms during the cosine and sine FTTPS in the case where (a) the control waveform $\Omega_n(t)$ produces consistent $+\pi$-rotations along $X_n$ (``same sign") and (b) the control waveform produces alternating  $+\pi$- and $-\pi$-rotations  along $X_n$ (``alternating sign") . Although the sequences shown in Fig.~\ref{pulsewidths} only have two pulses, the higher order FTTPS can be constructed from repetitions of these ``base" sequences. If the duration of the base sequences is $T_k\equiv T/k$,  $k$th-order cosine and sine FTTPS of duration $T$ are formed by $k$ repetitions.

Depending on the sequences applied to qubits 1 and 2, the choice of waveform (same vs. alternating sign) has a substantial effect on the contribution of  pulse width errors. Below, we examine the sequences applied to qubits 1 and 2 in the QNS protocol and the choice of waveforms that minimizes pulse width errors.\\

\emph{Case 1:}  Cosine FTTPS of order $k$ applied to qubit 1(2), free evolution applied to qubit 2(1). Consider first the scenario where a cosine FTTPS of order $k$ is applied to qubit 1 and free evolution is applied to qubit 2. In this case, $\sin[\theta_2(t)]=0$ and the contribution of pulse width errors increases with $\int_0^{T_k}dt \sin[\theta_1(t)]$, where $T_k$ is the time duration of the base sequences in Fig.~\ref{pulsewidths}. Note that for the same sign waveform, $\sin[\theta_1(t)]$ alternates sign so that $\int_0^{T_k}dt \sin[\theta_1(t)]\approx 0$, unlike the alternating sign waveform. A similar argument applies when cosine FTTPS of order $k$ are applied to qubit 2 and free evolution is applied to qubit 1. Therefore, when a single qubit is being controlled, applying $\pi$-pulses with the same sign minimizes pulse width errors.\\

\emph{Case 2:}  Cosine FTTPS of order $k$ applied to qubits 1 and 2 simultaneously. When a cosine FTTPS is applied to the qubits simultaneously, the contribution from pulse width error depends on $\int_0^{T_k}dt \sin[\theta_1(t)]$, $\int_0^{T_k}dt \sin[\theta_2(t)]$, and three crosstalk-generated integrals:
\begin{align}
&\int_0^{T_k}dt \sin[\theta_1(t)]\sin[\theta_2(t)] \notag\\
&\int_0^{T_k}dt \cos[\theta_1(t)]\sin[\theta_2(t)] \notag\\
&\int_0^{T_k}dt \sin[\theta_1(t)]\cos[\theta_2(t)] \notag
\end{align} 

Note that $\cos[\theta_1(t)]$ and $\cos[\theta_2(t)]$ change sign whenever $\sin[\theta_2(t)]$ and $\sin[\theta_1(t)]$ are respectively nonzero, implying that $\int_0^{T_k}dt \cos[\theta_1(t)]\sin[\theta_2(t)]\approx \int_0^{T_k}dt \sin[\theta_1(t)]\cos[\theta_2(t)]\approx 0$ for both alternating and same sign pulses. As discussed in \emph{Case 1}, the contributions from  $\int_0^{T_k}dt \sin[\theta_1(t)]$ and $\int_0^{T_k}dt \sin[\theta_2(t)]$ are suppressed when the pulses on both qubits have the same sign. When the pulses on one qubit have the same sign and the pulses on the other qubit alternate sign, $\sin[\theta_1(t)]\sin[\theta_2(t)]$ alternates sign.  The pulse width error contributions from $\int_0^{T_k}dt \sin[\theta_1(t)]\sin[\theta_2(t)]$ are, therefore, suppressed when the pulses on one qubit have the same sign and the pulses on the other qubit alternate sign. Whether pulse width errors are better suppressed overall for same sign pulses on both qubits or alternating sign pulses on one qubit and same sign pulses on the other depends on the magnitudes of $\beta_1(t)+\Delta_1$ and $\beta_2(t)+\Delta_2$ relative to $\beta_{12}(t)+J$. \\

\emph{Case 3:}  Cosine FTTPS of order $k$ applied to qubit 1 and Sine FTTPS of order $k$ applied to qubit 2 simultaneously. For this case, the contribution from pulse width error again depends on the same integrals as in Case 2. Note that  $\int_0^{T_k}dt \sin[\theta_1(t)]\sin[\theta_2(t)]\approx 0$, since $\sin[\theta_1(t)]$ and $\sin[\theta_2(t)]$ are never nonzero at the same time. Note that $\cos[\theta_1(t)]\sin[\theta_2(t)]$ and $\sin[\theta_1(t)]\cos[\theta_2(t)]$ alternate sign when pulses with alternating sign are applied to both qubits, implying that $\int_0^{T_k}dt \cos[\theta_1(t)]\sin[\theta_2(t)]\approx 0$ and $\int_0^{T_k}dt \sin[\theta_1(t)]\cos[\theta_2(t)]\approx 0$. However, the pulse width error contributions from  $\int_0^{T_k}dt \sin[\theta_1(t)]$ and $\int_0^{T_k}dt \sin[\theta_2(t)]$ are suppressed when the pulses on both qubits have the same sign. Whether same sign or alternating sign pulses on both qubits reduce the overall pulse error again depends on the magnitudes of $\beta_1(t)+\Delta_1$ and $\beta_2(t)+\Delta_2$ relative to $\beta_{12}(t)+J$. 

\section{Estimating the static noise components and crosstalk}\label{app:detuning-crosstalk}

\begin{figure}[t]
    \centering
    \includegraphics[width=\columnwidth]{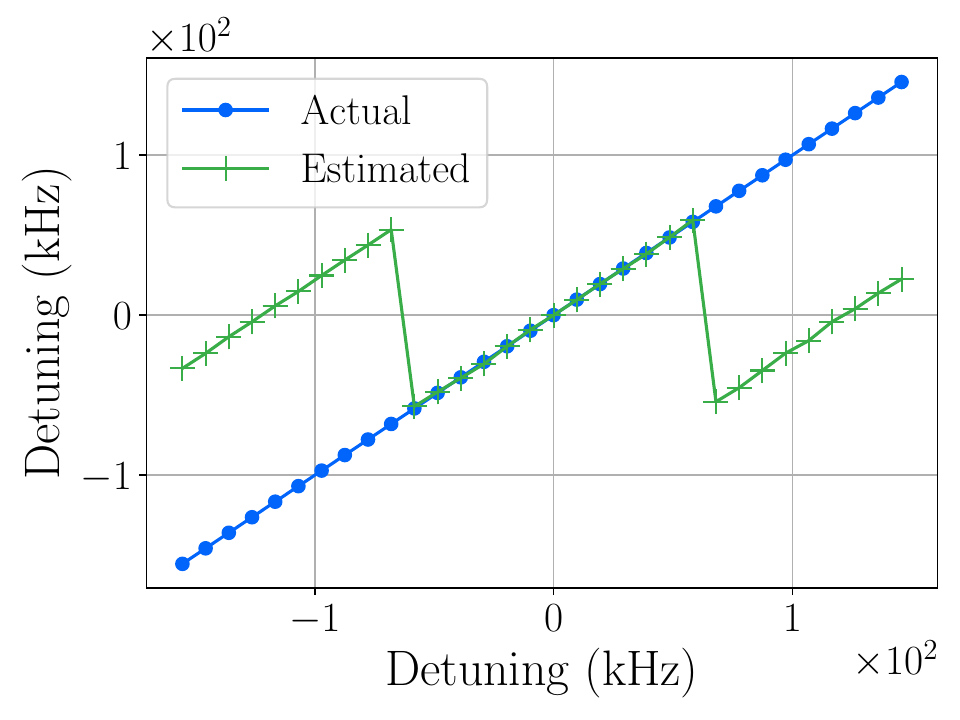}
    \caption{The detuning estimated from simulation and the actual detuning values used in the simulation. When the detuning value is very large it cannot be accurately estimated.}
    \label{fig:det}
\end{figure}

\begin{figure}[t]
    \centering
    \includegraphics[width=\columnwidth]{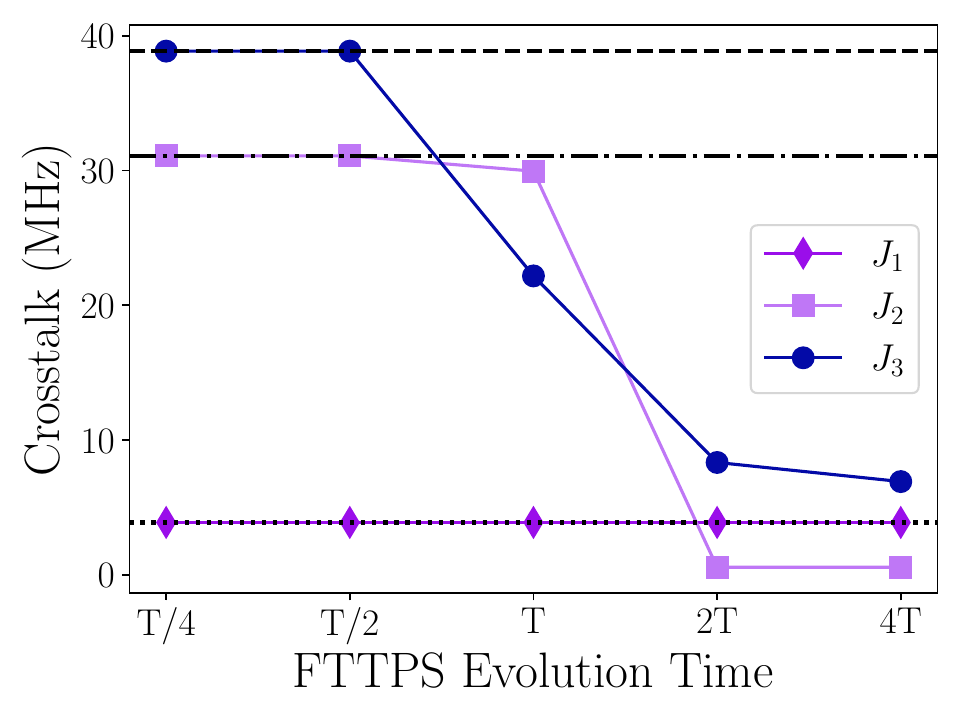}
    \caption{The estimated crosstalk as a function of total time $T$ for three different crosstalk values $J_{n}$. For small values, the crosstalk can be accurately estimated for any total time. As $J$ increases, shorter total times are needed to estimate the crosstalk.}
    \label{fig:xtalk}
\end{figure}

By combining targeted control, state preparations, and observables, we can also estimate the magnitudes of the static noise components and crosstalk in the weak noise limit. Similar to the decay constants, we can isolate the dynamical contributions of the static noise components and crosstalk. For $n\in\{1,2\}$,
\begin{align*}
\Theta_{n}(T)&=\tan^{-1}\left(\frac{\mathbb{E}_{XX}[Y_n(T)]}{\mathbb{E}_{XX}[X_n(T)]}\right),\\
\Theta_{12}(T)&=\cos^{-1}\left(\frac{A_n^{XX}(T)}{A_n^{XZ}(T)}\right)\\
\Theta_{n}(T)+\Theta_{12}(T)&=\tan^{-1}\left(\frac{\mathbb{E}_{XZ}[Y_n(T)]}{\mathbb{E}_{XZ}[X_n(T)]}\right),
\end{align*}
where
\begin{align*}
A_n^{XX}(T)^2 &\equiv \mathbb{E}_{XX}[X_n(T)]^2+\mathbb{E}_{XX}[Y_n(T)]^2,\\
A_n^{XZ}(T)^2 &\equiv \mathbb{E}_{XZ}[X_n(T)]^2+\mathbb{E}_{XZ}[Y_n(T)]^2.
\end{align*}
It follows from Eqs.~\eqref{eq:C1Obs1} and \eqref{eq:C1Obs2} that when free evolution or, equivalently, the cosine FTTPS of order $k$ is applied to both qubits
\begin{align*}
\Theta_{n}(T)&=2\Delta_nT,\\
\Theta_{12}(T)&=2JT.
\end{align*}
These expressions enable us to solve for $\Delta_n$ and $J$ from measured values of the dynamical contributions. However, as discussed in Section~\ref{sec:Simulations}, detuning and crosstalk must meet the following conditions:   $-\pi\leq2T\Delta_{n}\leq\pi$ and $0\leq2TJ\leq2\pi$. When the values are outside this regime the phase will wrap. In Fig.~\ref{fig:det}, we simulate the actual detuning and estimated detuning for $T=4.096\mu$s for $q_{1}$, with the detuning of $q_{2}$ and crosstalk held constant. We observe phase wrapping, which results in an error in the estimation of the detuning frequency. One solution would be to use \acrshort{fttps} with shorter total time. Since the estimates only require the first FTTP sequence, it adds an overhead of six extra sequences. In Fig.~\ref{fig:xtalk}, we compare the estimates of three crosstalk values as a function of \acrshort{fttps} evolution time. For very small $J$, the crosstalk is accurately estimated. As the crosstalk increases, shorter evolution times are needed to accurately estimate the static crosstalk strength. Once the phase wrapping condition is violated, the value will be too large and other methods, like joint amplification of $ZZ$~\cite{takita2017experimental:jazz}, will be needed to estimate the crosstalk, or Ramsey for the detuning.

\section{Linear Inversion}\label{linarinv}

Here, we summarize the necessary components of the linear inversion procedure used to perform spectrum reconstruction. The sequences are defined in Sec.~\ref{subsec:sum-protocol}, with the inversion problem defined in Sec.~\ref{sec:ControlSec}. Here, we will focus on the relationship between the decay rates and the FFs resulting from the FTTPS protocol.

Let $G_{n,n'}^{(i,k_1,k_2)}(\omega,T)$ with $(n,n')\in\{(1,1),$ $(2,2),$ $(1,2),$ $(12,12)\}$ be a filter function generated by sequence combination $i\in\{1,2,3,4\}$ when an FTTPS of order $k_1$ is applied to qubit 1 and an FTTPS of order $k_2$ is applied to qubit 2.  Similarly, let $[\chi(T)]^{(i,k_1,k_2)}$ denote a decay rate experimentally measured after applying combination $i$ to the qubits when an FTTPS of order $k_1$ is applied to qubit 1 and an FTTPS of order $k_2$ is applied to qubit 2.

To define the linear system that we use to estimate the spectra, we discretize the angular frequency axis into increments $\Delta\omega_0\equiv[-\pi/T,\pi/T]$, $\Delta\omega_1\equiv[\pi/T,3\pi/T],\,\ldots$, $\Delta\omega_K\equiv[(2\pi K-1)/T,(2\pi K+1)/T]$, where $K$ is the maximum FTTPS order used. We define the following $K$-dimensional vectors consisting of the FFs and spectra discretized over the frequency increments,
\begin{widetext}
\begin{align*}
&\vec{G}_{n,n'}^{(i,k_1,k_2)}=\left[\int_{\Delta\omega_0}d\omega\,G_{n,n'}^{(i,k_1,k_2)}(\omega,T),\; \int_{\Delta\omega_1}d\omega\,G_{n,n'}^{(i,k_1,k_2)}(\omega,T),...,\int_{\Delta\omega_K}d\omega\,G_{n,n'}^{(i,k_1,k_2)}(\omega,T)\right],\\
&\vec{S}_{n,n'}=[{S}_{n,n'}(0),{S}_{n,n'}(2\pi/T),\ldots,{S}_{n,n'}(2\pi K/T)].
\end{align*}
\end{widetext}

The FTTPS protocol results in the following set of linear equations for each combination outlined in Sec.~\ref{subsec:sum-protocol}. For sequence combinations 1 and 2, which include free evolution on one qubit and cosine FTTPS on the other, the decay rate relationships are given by
\begin{align*}
&[\chi_{1,1;12,12}(T)]^{(1,k,0)}=\frac{2}{\pi}\vec{G}_{1,1}^{(1,k,0)}\cdot\vec{S}_{1,1}+\frac{2}{\pi}\vec{G}_{12,12}^{(1,k,0)}\cdot\vec{S}_{12,12}\\
&[\chi_{2,2;12,12}(T)]^{(2,0,k)}=\frac{2}{\pi}\vec{G}_{2,2}^{(2,0,k)}\cdot\vec{S}_{2,2}+\frac{2}{\pi}\vec{G}_{12,12}^{(2,0,k)}\cdot\vec{S}_{12,12}.
\end{align*}
For the case where cosine FTTPS is applied on both qubits, the set of linear equations is
\begin{align*}
[\chi_{1,1;2,2}(T)]^{(3,k,k)}&=\frac{2}{\pi}\vec{G}_{1,1}^{(3,k,k)}\cdot\vec{S}_{1,1}+\frac{2}{\pi}\vec{G}_{2,2}^{(3,K,K)}\cdot\vec{S}_{2,2}\\
[\chi_{12}(T)]^{(3,k,k)}&=\frac{4}{\pi}\text{Re}[\vec{G}_{1,2}^{(3,k,k)}]\cdot\text{Re}[\vec{S}_{12}].
\end{align*}
Finally, for the case of cosine FTTPS applied to one qubit and sine FTTPS applied to the other:
\begin{align*}
[\chi_{1,1;2,2}(T)]^{(4,k,k)}&=\frac{2}{\pi}\vec{G}_{1,1}^{(4,k,k)}\cdot\vec{S}_{11}+\frac{2}{\pi}\vec{G}_{2,2}^{(4,k,k)}\cdot\vec{S}_{22}\\
[\chi_{12}(T)]^{(4,k,k)}&=-\frac{4}{\pi}\text{Im}[\vec{G}_{1,2}^{(4,k,k)}]\cdot\text{Im}[\vec{S}_{12}].
\end{align*}
In all scenarios, $k=0,\ldots, K-1$, yielding $7K$ total equations to solve to obtain estimates for the four spectra.

\section{State preparations, sequences, and measurements}\label{sequences}

To obtain the required noise coefficients experimentally, we use the following state preparations, sequences, and measurements. Assume $k\in\{1,...K-1\}$ unless stated otherwise. 

\begin{enumerate}
    \item Prepare qubits in $|+X\rangle_{1}|+Z\rangle_{2}$.
    \begin{enumerate}
        \item Apply free evolution to both qubit 1 and qubit 2. Measure $X_{1}$.
        \item Apply free evolution to both qubit 1 and qubit 2. Measure $Y_{1}$.
        \item Apply cosine \acrshort{fttps} of order $k$ to qubit 1 and free evolution to qubit 2. Measure $X_{1}$.
    \end{enumerate}
    \item Prepare qubits in $|+Z\rangle_{1}|+X\rangle_{2}$.
    \begin{enumerate}
        \item Apply free evolution to both qubit 1 and qubit 2. Measure $X_{2}$.
        \item Apply free evolution to both qubit 1 and qubit 2. Measure $Y_{2}$.
        \item Apply cosine \acrshort{fttps} of order $k$ to qubit 2 and free evolution to qubit 1. Measure $X_{2}$.
    \end{enumerate}
    \item Prepare qubits in $|+X\rangle_{1}|+X\rangle_{2}$.
    \begin{enumerate}
        \item Apply free evolution to both qubit 1 and qubit 2. Measure $X_{1}$ and $X_{2}$.
        \item Apply free evolution to both qubit 1 and qubit 2. Measure $Y_{1}$ and $Y_{2}$.
        \item Apply free evolution to both qubit 1 and qubit 2. Measure $X_{1}$ and $Y_{2}$.
        \item Apply free evolution to both qubit 1 and qubit 2. Measure $Y_{1}$ and $X_{2}$.
        \item Apply cosine \acrshort{fttps} of order $k$ to both qubit 1 and qubit 2. Measure $X_{1}$ and $X_{2}$.
        \item Apply cosine \acrshort{fttps} of order $k$ to both qubit 1 and qubit 2. Measure $Y_{1}$ and $Y_{2}$.
    \end{enumerate}
    \item Prepare qubits in $|+X\rangle_{1}|+X\rangle_{2}$.
    \begin{enumerate}
        \item Apply cosine \acrshort{fttps} of order $k$ to qubit 1 and sine \acrshort{fttps} of order $k$ to qubit 2. Measure $X_{1}$ and $X_{2}$.
        \item Apply cosine \acrshort{fttps} of order $k$ to qubit 1 and sine \acrshort{fttps} of order $k$ to qubit 2. Measure $Y_{1}$ and $Y_{2}$.
    \end{enumerate}
\end{enumerate}

\section{SchWARMA-Based Numerical Simulation Framework}
\label{app:schwarma}
The numerical simulations were performed by discretizing the evolution according to
\begin{equation}
U(\bm{\Phi}) = E(\Phi_N) G_N  E(\Phi_{N-1}) G_{N-1} \cdots  E(\Phi_1)G_1.
\end{equation}
The unitary $G_j=G^{(1)}_{j} \otimes G^{(2)}_{j}$ is composed of unitaries $G^{(k)}_j$ which denote the $j$th gate applied to qubit $k$ in the FTTPS evolution. Noise enters the evolution via the parametrized error unitary $E(\Phi_j)$. In the case of local dephasing on both qubits and quantum crosstalk, $\Phi_j=(\phi^{(1)}_j, \phi^{(2)}_j, \phi^{(zz)}_j)$ and the error unitary can be explicitly expressed as $E(\Phi_j) = R_{z_1}(\phi^{(1)}_j) R_{z_2}(\phi^{(2)}_j) R_{zz}(\phi^{(zz)}_j)$. The gate $R_{z_k}(\phi^{(k)})$ represents a $z$-rotation with time correlated angle $\phi^{(k)}_j$ for qubit $k$, while $R_{zz}(\phi)$ denotes a rotation generated by $Z_1Z_2$. Upon specifying the noise mean and power spectral density, a SchWARMA model can be used to generate a trajectory $\bm{\Phi}$. Averaging expectation values over numerous trajectories results in the desired dynamics.

In the SchWARMA model approach, temporal correlations are created by driving an ARMA model with white Gaussian noise. The ARMA model is formally defined as 
\begin{equation}
y_k = \sum^{p}_{i=1} a_i y_{i-k} + \sum^{q}_{j=0} b_j w_{k-j},
\end{equation}
where the coefficients $\{a_i\}$ dictate the autoregressive components, while ${b_j}$ define the moving average coefficients. The variable $w_k$ denotes a white Gaussian input. This filtering approach yields a power spectrum 
\begin{equation}
S_y(\omega) = \frac{| \sum^{q}_{k=0} b_k \exp(-i k \omega ) |^2}{| 1 + \sum^{p}_{k=1} a_k \exp(-i k \omega ) |^2}.
\end{equation}
It is known that ARMA models can approximate any discrete-time power spectrum to arbitrary accuracy~\cite{holan2010arma}.  Here, we use ARMA models to define the OU process. In particular, it is known that the OU process can be defined as an AR1 model with ARMA coefficients $a_1 = 1$, $a_2 = -\zeta_1$, $b_0=\sqrt{\zeta_2}$, with $\zeta_1 = e^{-\theta \delta t}$ and $\zeta_2 = \sigma^2 (1-\zeta^2_1)/(2\theta) $. All other ARMA parameters are identically zero. We use this ARMA parameterization to construct each of the $\eta(t)$ required for the local spatiotemporally correlated dephasing and quantum crosstalk noise.

\bibliography{refs}

\end{document}